\documentclass[aps,prl,twocolumn,superscriptaddress,showpacs,amsmath,amssymb]{revtex4-1}

\usepackage{graphicx,bbm}
\usepackage{hyperref}
\usepackage{mathtools}
\usepackage{color}
\usepackage{braket}
\usepackage{bm}
\usepackage{tikz}

\def\cA{{\mathfrak a}}    
\def\cB{{\mathfrak b}}    
\def\cC{{\mathfrak c}} 
\def\cE{{\vartheta}} 

\def\cO{{\mathcal O}}

\def \LL {{\cal L}}

\newcommand{\er}[1]{Eq.~\eqref{#1}}

\begin{document}

\title{Exact large deviation statistics and trajectory phase transition of a deterministic boundary driven cellular automaton}

\author{Berislav Bu\v{c}a}
\affiliation{Clarendon Laboratory, University of Oxford, Parks Road, Oxford OX1 3PU, United Kingdom.}
\author{Juan P. Garrahan}
\affiliation{School of Physics and Astronomy 
and
Centre for the Mathematics and Theoretical Physics of Quantum Non-equilibrium Systems,
University of Nottingham, Nottingham NG7 2RD, United Kingdom}
\author{Toma\v{z} Prosen}
\author{Matthieu Vanicat}
\affiliation{Faculty of Mathematics and Physics, University of Ljubljana, Jadranska 19, SI-1000 Ljubljana, Slovenia.}

\begin{abstract}
We study the statistical properties of the long-time dynamics of the rule 54 reversible cellular automaton (CA), driven stochastically at its boundaries. This CA can be considered as a discrete-time and deterministic version of the Fredrickson-Andersen kinetically constrained model (KCM). By means of a matrix product ansatz, we compute the exact large deviation cumulant generating functions for a wide range of time-extensive observables of the dynamics, together with their associated rate functions and conditioned long-time distributions over configurations. We show that for all instances of boundary driving the CA dynamics occurs at the point of phase coexistence between competing active and inactive dynamical phases, similar to what happens in more standard KCMs. We also find the exact finite size scaling behaviour of these trajectory transitions, and provide the explicit ``Doob-transformed'' dynamics that optimally realises rare dynamical events. 
\end{abstract}

\maketitle

\noindent
{\bf \em Introduction.--} Classical systems which evolve stochastically subject to constraints display complex dynamics, often beyond what can be anticipated simply from their static properties. This is what occurs in the presence of excluded volume interactions, such as in simple exclusion processes \cite{Derrida2007,Mallick2015}, or when configuration space is restricted, such as in dimer coverings \cite{Henley2010,Chalker2017}, or in systems where dynamical rules are subject to {\em kinetic constraints}, as for example in kinetically constrained models (KCMs) of glasses \cite{Ritort2003,Garrahan2011}. Constrained dynamics is also proving increasingly relevant to quantum many-body systems, including problems such as slow thermalisation and non-ergodicity in the absence of disorder \cite{Horssen2015,Smith2017,Shiraishi2017,Lan2018,Turner2018}, operator spreading and entanglement growth \cite{Nahum2017,Keyserlingk2018,Rowlands2018,Chen2018,Gopalakrishnan2018,Knap2018,Tran2018,Gopalakrishnan2018b}, and in the dynamics of ensembles of Rydberg atoms \cite{Lesanovsky2013,Urvoy2015,Valado2016}. 

Complex collective dynamics must be characterised through the statistical properties of dynamical observables, something which can be readily done by means of large deviation (LD) techniques \cite{Lecomte2007,Garrahan2007,Garrahan2009,Touchette2009}. This allows to study ensembles of trajectories of the dynamics as one would study ensembles of configurations in equilibrium statistical mechanics. Among other things, the dynamical LD approach reveals in many systems the existence of competing {\em dynamical phases} and the corresponding phase transitions between them, as for example in KCMs \cite{Garrahan2007,Garrahan2009}, exclusion processes \cite{Appert-Rolland2008,Espigares2013,Jack2015,Karevski2017}, dimer models \cite{Oakes2018}, and several other classical \cite{Hedges2009,Speck2012,Weber2013,Baek2017} and quantum \cite{Garrahan2010} systems. This rich phase structure of trajectory space is what underlies the complex dynamics of these systems. 

Here we generalise the above ideas to systems whose (bulk) dynamics is {\em deterministic} and {\em reversible}. We consider specifically the ``rule 54'' cellular automaton (CA) of Ref.~ \cite{Bobenko1993} (RCA54). The local rules that define the interactions of this CA (see below) are similar to the kinetic constraints of the simplest of KCMs, the (one-spin facilitated) Fredrickson-Andersen (FA) model \cite{Fredrickson1984,Ritort2003,Garrahan2018}. As such the RCA54 is referred to also as the ``Floquet-FA'' model \cite{Gopalakrishnan2018} since it can be considered a synchronous, discrete and deterministic version of the FA model. (The RCA54 is also related to the ERCA 250R of Takesue \cite{Takesue1987}.) A remarkable property of the RCA54 is that it is {\em integrable} \cite{Prosen2016} and, in presence of stochastic driving at its boundaries, one can obtain exactly its (in general non-equilibrium) steady state distribution \cite{Prosen2016,Inoue2018}, certain decay modes \cite{Prosen2017}, and dynamical structure factors \cite{Klobas2018} in terms of matrix product states.  

In this paper we compute the {\em exact large deviation statistics} of the boundary driven RCA54 by generalising the methods of Refs.\ \cite{Prosen2016,Inoue2018,Prosen2017,Klobas2018}. Via a novel inhomogeneous matrix product ansatz we obtain the exact  cumulant generating functions and rate functions of a broad class of time-extensive observables of the dynamics. We prove the existence of distinct active and inactive dynamical phases, with the dynamics of the RCA54 occurring at the phase transition point. To our knowledge, our findings here represent the only exact results for LDs in interacting models beyond those for simple exclusion processes \cite{Derrida1998,Appert-Rolland2008,Prolhac2010,Gier2011,Gorissen2012,Crampe2016}, and the first for bulk-deterministic systems. 

\smallskip

\noindent
{\bf \em Model.--} We consider a system defined by binary variables $n_i\in\{0,1\}$ (up/down state) on sites $i\in\{1,\ldots,N\}$ of a lattice with even size $N$. A configuration at time $t$ is described by a binary string $\bm{n}^t=(n_1^t,n_2^t,\dots,n_N^t)$.
The dynamics in the bulk is given by the discrete, deterministic RCA54 \cite{Bobenko1993}, while the dynamics on the boundary sites is stochastic \cite{Prosen2016,Prosen2017}.

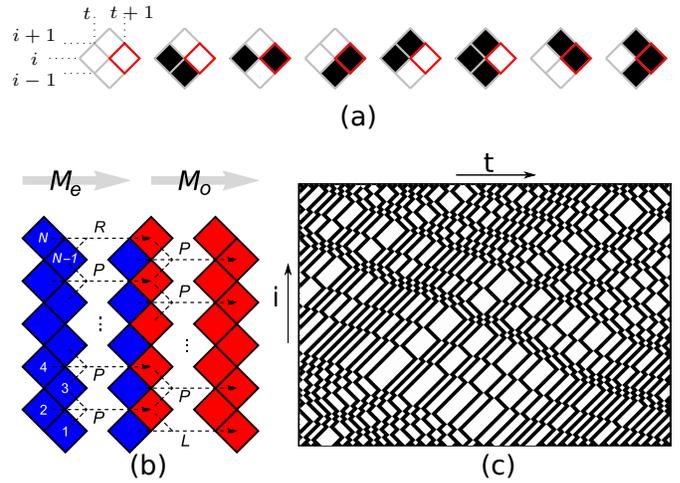
\begin{figure}[htb]

\begin{center}
 \hspace{-1cm}
 \begin{tikzpicture}[scale=0.4]
\foreach \i in {2.5,7.5,...,17.5}
{\draw[fill] (\i+0.5,-0.5) -- (\i+1,0) -- (\i+1.5,-0.5) -- (\i+1,-1) -- cycle ;}
\foreach \i in {10,12.5,...,17.5}
{\draw[fill] (\i+0.5,0.5) -- (\i+1,1) -- (\i+1.5,0.5) -- (\i+1,0) -- cycle ;}
\foreach \i in {5,7.5,15,17.5}
{\draw[fill] (\i+1,0) -- (\i+1.5,0.5) -- (\i+2,0) -- (\i+1.5,-0.5) -- cycle ;}
\foreach \i in {2.5,5,10,12.5}
{\draw[fill] (\i,0) -- (\i+0.5,0.5) -- (\i+1,0) -- (\i+0.5,-0.5) -- cycle ;}
\foreach \i in {0,2.5,...,17.5}
{\draw[thick,lightgray] (\i,0) -- (\i+1,1) ; \draw[thick,lightgray] (\i,0) -- (\i+1,-1) ; \draw[thick,lightgray] (\i+1,1) -- (\i+2,0) ; \draw[thick,lightgray] (\i+1,-1) -- (\i+2,0) ;
\draw[thick,lightgray] (\i+0.5,0.5) -- (\i+1.5,-0.5) ; \draw[thick,lightgray] (\i+0.5,-0.5) -- (\i+1.5,0.5) ;
\draw[thick,red] (\i+1,0) -- (\i+1+0.5,0.5) -- (\i+1+1,0) -- (\i+1+0.5,-0.5) -- cycle ;}
\draw[dotted] (-1,0) -- (0,0);
\draw[dotted] (-0.5,-0.5) -- (0.5,-0.5);
\draw[dotted] (-0.5,0.5) -- (0.5,0.5);
\draw[dotted] (0.5,0.5) -- (0.5,1.5);
\draw[dotted] (1.5,0.5) -- (1.5,1.5);
\node at (-1.5,0) [] {\begin{scriptsize}$i$\end{scriptsize}}; 
\node at (-1.5,-0.75) [] {\begin{scriptsize}$i-1$\end{scriptsize}};
\node at (-1.5,0.75) [] {\begin{scriptsize}$i+1$\end{scriptsize}};
\node at (0.25,1.5) [] {\begin{scriptsize}$t$\end{scriptsize}};
\node at (1.75,1.5) [] {\begin{scriptsize}$t+1$\end{scriptsize}};
\draw[dotted,white] (-4,0) -- (-3,0);
 \end{tikzpicture}
 \end{center}
 \begin{center}
 \vspace{-3mm}
\includegraphics[width=\columnwidth]{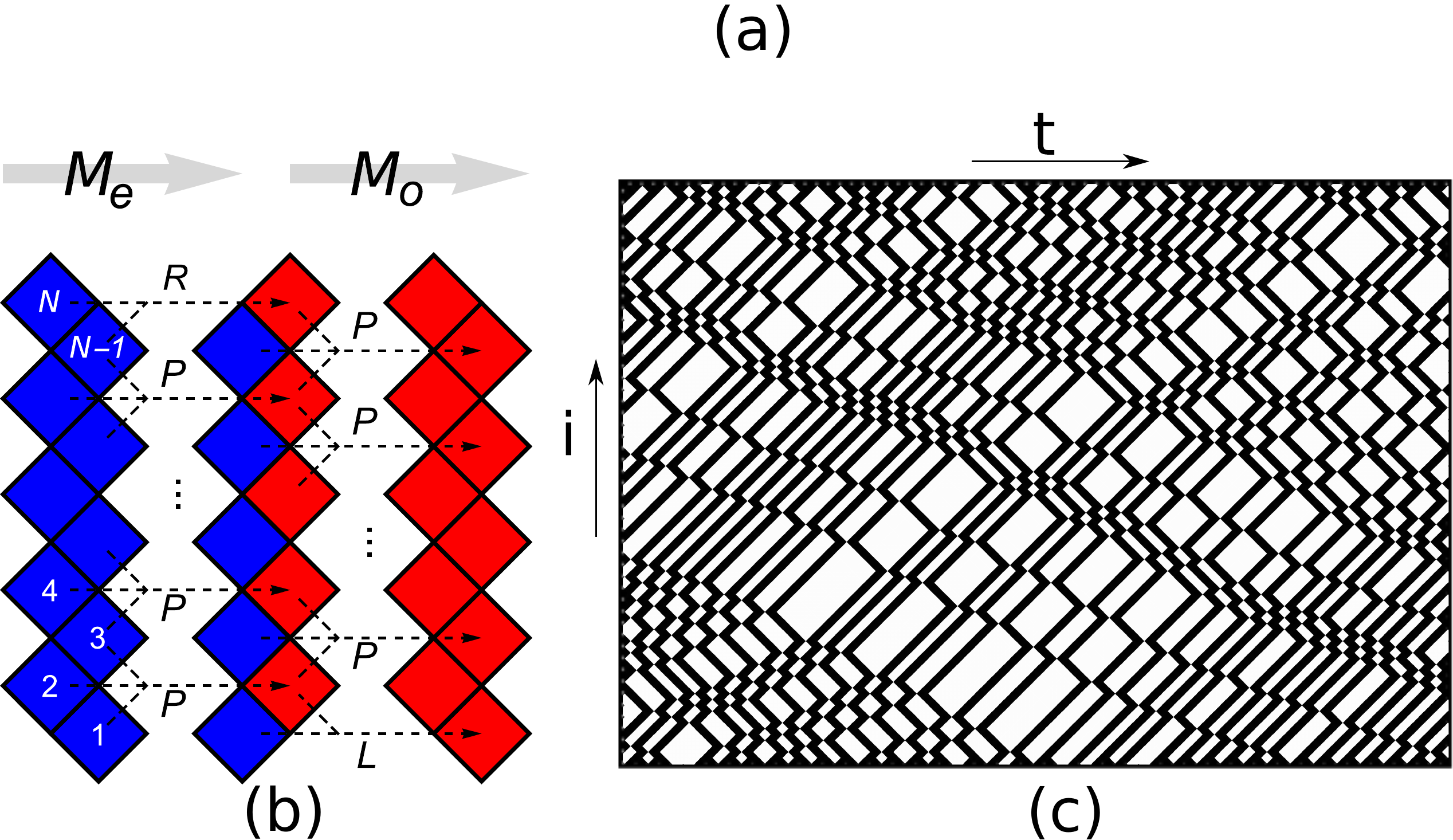}
\end{center}
\vspace{-3mm}

\caption{Boundary-driven RCA54: (a) Deterministic local dynamical rules for bulk dynamics. (b) Action of the propagator in the two half-time steps. (c) A typical trajectory of the model for $(N,T)=(100,75)$ and stochastic boundaries with $(\alpha,\beta,\gamma,\delta)=(1/3,1/8,1/2,2/5)$. 
} 
\label{fig:def_dynamic}
\end{figure}

The update rule is decomposed into two Floquet-like half-time steps. During the first half-time step, $\bm{n}^t \to \bm{n}^{t+1/2}$, only even sites are updated, so that $n_i^{t+1/2} = n_i^t$ for $i$ odd. For all $i$ even with $2\leq i\leq N-2$, the evolution of $n_i^t$ is deterministic through the relation $n_i^{t+1/2}=\chi(n_{i-1}^t,n_i^t,n_{i+1}^t)$, where $\chi(n,n',n'')=n+n'+n''+n n'' \mod 2$ is the rule-54 function \cite{Bobenko1993}, see Fig.\ \ref{fig:def_dynamic}(a). This local update rule is similar to the constraint of the FA model: a site can flip only if at least one of its nearest neighbours is in the up state \cite{Fredrickson1984}. The last site is updated stochastically depending on the state of its neighbour: $n_N^{t+1/2} = 0$ with probability $\gamma + n_{N-1}^t (\delta - \gamma)$ or $n_N^{t+1/2} = 1$ otherwise. In the second half-time step,
$\bm{n}^{t+1/2} \to \bm{n}^{t+1}$, for even sites we have $n_i^{t+1} = n_i^{t+1/2}$, while odd sites in the bulk evolve deterministically with rule-54, $n_i^{t+1}=\chi(n_{i-1}^{t+1/2},n_i^{t+1/2},n_{i+1}^{t+1/2})$. The first site is updated stochastically, with $n_1^{t+1} = 0$ with probability $\alpha + n_{2}^{t+1/2} (\beta - \alpha)$ or $n_1^{t+1} = 1$ otherwise. 

The above rules define a discrete-time, irreducible and non-reversible Markov process. Physically it models a gas of solitons stochastically
emitted from reservoirs at the boundaries, propagating at constant unit velocity in the bulk and interacting pairwise through a one-time step delay \cite{Prosen2016,Prosen2017,Inoue2018}. Depending on the boundary rates, the system is driven out-of-equilibrium by the reservoirs leading to a net flow of solitons in the stationary state.

We define $p_{\bm{n}}^t$ to be the probability that $\bm{n}^t=\bm{n}$ and 
$\bm{p}^t= \sum_{n_1,n_2,\dots,n_N\in \{0,1\}}p_{\bm{n}}^t \ e_{n_1} \otimes e_{n_2} \otimes \dots \otimes e_{n_N}$ the associated probability vector in $(\mathbb{R}^2)^{\otimes N}$ (where $e_0$ and $e_1$ denote
the elementary basis of $\mathbb{R}^2$). The master equation can be written as
$\bm{p}^{t+1}=M\bm{p}^t$ where the Markov matrix $M = M_{\rm o}M_{\rm e}$ is expressed as the product of two operators associated with the even and odd half-time steps, see Fig.\ \ref{fig:def_dynamic}(b), 
\begin{eqnarray}
  M_{\rm e} & = & P_{123}P_{345}\cdots P_{N-3,N-2,N-1} R_{N-1,N}  \nonumber \\
  M_{\rm o} & = & L_{12} P_{234}P_{456} \cdots P_{N-2,N-1,N}.
\end{eqnarray}
 The subscripts indicate on which sites of the lattice the operators are acting non-trivially. The operator $P$ is the $8\times 8$ permutation matrix (acting on three sites) that enforces the 
dynamical rule in the bulk, with elements $P_{nn'n'\!'}^{mm'm'\!'}= \delta_{n,m}\delta_{\chi(nn'n'\!'),m'}\delta_{n'\!',m'\!'}$ where $\delta_{n,m}$ is the Kronecker symbol. The operators $L$ and $R$ are the $4\times 4$ stochastic matrices for the boundary processes of first and last site, respectively. A typical trajectory of the RCA54 is shown in Fig.\ \ref{fig:def_dynamic}(c). For further details of the model see \cite{Prosen2016,Prosen2017,Inoue2018} and \cite{SM}.

\smallskip

\noindent
{\bf \em Large deviations of time-integrated observables.--} We are interested in the statistics of general (possibly inhomogeneous) space- and time-extensive observables of the form 
\begin{equation}
 \cO_T = \sum_{t=0}^{T-1} \sum_{j=1}^{N-1}\Big[f_j(n^t_j,n^t_{j+1})+g_j(n^{t+1/2}_j,n^{t+1/2}_{j+1})\Big]
 \label{obs}
\end{equation}
in the large time $T$ limit. These are dynamical (or trajectory) observables as they depend on the full time history 
$(\bm{n}^0,\bm{n}^{1/2},\bm{n}^1,\dots,\bm{n}^{T-1/2})$. An example is the time-integrated number of up sites (which is not conserved in the RCA54) corresponding to $f_j(n,n') = (n+n')/2$ and $g_j(n,n')=0$.

For large $T$ the probability of $\cO_T$ has a LD form, $P_T(\cO) = \langle \delta( \cO - \cO_T)\rangle \sim_{T \to \infty} e^{-T \varphi_N(\cO_T / T)}$, where $\varphi_N(x)$ is the {\em rate function} (where the subscript indicates its size-dependence). 
The moment generating function also has a LD form, $Z_T(s) = \langle e^{-s \cO_T} \rangle \sim e^{T \theta_N(s)}$, where $\theta_N(s)$ is called the {\em scaled cumulant generating function} (SCGF),
as its derivatives at $s=0$ correspond to the cumulants of $\cO_T$ divided by time. 
The LD functions play the role of free-energies for trajectories and are related by a Legendre transform, $\theta_N(s) = - \min_x \left[ s x + \varphi_N(x) \right]$. 

To obtain the SCGF we deform, or {\em tilt}, the Markov matrix \cite{Touchette2009}: we define
$M(s) = M_{\rm o} \, G(s) \, M_{\rm e} \, F(s)$, where we have introduced the diagonal operators 
$F_{\bm{n},\bm{n'}}(s) = \delta_{\bm{n},\bm{n'}} \prod_{i=1}^{N-1} f_{n_i,n_{i+1}}^{(i)}$ and
$G_{\bm{n},\bm{n'}}(s) = \delta_{\bm{n},\bm{n'}} \prod_{i=1}^{N-1} g_{n_i,n_{i+1}}^{(i)}$,
with the shorthand notation $f_{n,n'}^{(i)}=e^{-s f_i(n,n')}$ and $g_{n,n'}^{(i)}=e^{-s g_i(n,n')}$. We then have that $\theta_N(s) = \ln \lambda(s)$,   
where $\lambda(s)$ is the largest real eigenvalue of $M(s)$.

\smallskip

\noindent
{\bf \em Exact results from Matrix Ansatz.--} 
During the last decades, a technique called Matrix Ansatz has proven to be very efficient for deriving exact results in out of equilibrium systems. It has been introduced to compute analytically the stationary state of Markov chains \cite{Derrida1993} 
(see also \cite{Vanicat2017,Vanicat2018} for recent developments) and has later on been used to compute deformed ground state \cite{Gorissen2012,Finn2017}, 
eigenvectors \cite{Crampe2011,Prosen2017} or time-evolution of 
particular initial states \cite{Klobas2018}. We use it here to compute the ground-state of the tilted Markov operator $M(s)$.
Following the approach of \cite{Prosen2016,Prosen2017}, our strategy is to look for vectors $\bm{p}$ and $\bm{p'}$ such that
  $M_{\rm e}F(s) \bm{p} = \lambda_{R}(s) \bm{p'}$ and $M_{\rm o}G(s) \bm{p'} = \lambda_{L}(s) \bm{p}$. It then follows that $\lambda(s) = \lambda_{R}(s) \lambda_{L}(s)$ is the dominant eigenvalue of the tilted Markov operator, $M(s) \bm{p} = \lambda(s) \bm{p}$. 

It turns out that one can construct four pairs of site-dependent $3\times 3$ matrices
$W_n^{(j)}$, $V_n^{(j)}$, $X_n^{(j)}$, $Y_n^{(j)}$ \cite{SM} which satisfy the inhomogeneous bulk relations, for $j$ even:
\begin{eqnarray}
& f^{(j-1)}_{nn'} f^{(j)}_{n'n'\!'} W^{(j-1)}_n W^{(j)}_{n'} X^{(j+1)}_{n'\!'} = 
X^{(j-1)}_{n} V^{(j)}_{\chi(nn'n'\!')} V^{(j+1)}_{n'\!'}, \nonumber \\
& g^{(j-2)}_{nn'} g^{(j-1)}_{n'n'\!'} X^{(j-2)}_n V^{(j-1)}_{n'} V^{(j)}_{n'\!'} = W^{(j-2)}_n W^{(j-1)}_{\chi(nn'n'\!')} X^{(j)}_{n'\!'},  \nonumber
\end{eqnarray}
as well as six row 3-vectors $\bra{l_n},\bra{l'_{nn'}}$ and six column 3-vectors $\ket{r_{nn'}}$, $\ket{r'_n}$, satisfying the boundary equations
\begin{align*}
f^{(1)}_{nn'}f^{(2)}_{n'n'\!'} \bra{l_n}W^{(2)}_{n'} X^{(3)}_{n'\!'} & = \bra{l'_{n\chi(nn'n'\!')}}V^{(3)}_{n'\!'},  \\
\sum_{m,m'=0,1} R_{nn'}^{mm'}f^{(N-1)}_{mm'}\ket{r_{mm'}} & = \lambda_{\rm R} X^{(N-1)}_{n}\ket{r'_{n'}}, \\
\sum_{m,m'=0,1} L_{nn'}^{mm'} g^{(1)}_{mm'} \bra{l'_{mm'}} & = \lambda_{\rm L} \bra{l_n}X^{(2)}_{n'},\\
g^{(N-2)}_{nn'}\!g^{(N-1)}_{n'n'\!'} X^{(N-2)}_{n} V^{(N-1)}_{n'}\!\ket{r'_{n'\!'}} & = W^{(N-2)}_{n}\!\ket{r_{\!\chi(nn'n'\!')n'\!'}},
\end{align*}
These equations provide a cancellation scheme implying that an eigenvector of $M$, specifically vectors $\boldsymbol p$ and $\boldsymbol p'$, take the matrix product form
\begin{eqnarray}
  p_{n_1,\dots,n_N} &=&  \bra{l_{n_1}} W_{n_2}^{(2)}W_{n_3}^{(3)} \cdots W_{n_{N-3}}^{(N-3)}W_{n_{N-2}}^{(N-2)}\ket{r_{n_{N-1}n_N}} \nonumber \\
  p'_{n_1,\dots,n_N} &=&  \bra{l'_{n_1 n_2}} V_{n_3}^{(3)}V_{n_4}^{(4)} \cdots V_{n_{N-2}}^{(N-2)}V_{n_{N-1}}^{(N-1)}\ket{r'_{n_N}},
  \end{eqnarray}
 An explicit expression of the matrices and boundary vectors in the 3-dimensional auxilliary space are provided in \cite{SM}.
 The eigenvalue $\lambda=\lambda_{\rm L}\lambda_{\rm R}$ is proven to be the dominant root of a polynomial of order 4
 \begin{equation}
\lambda^4 - \alpha \gamma \cA_N \lambda^3 - \omega \cA_N^2 \lambda^2 - \beta\delta\xi \cA_N^3 \lambda + \eta \cA_N^4=0,
\label{lambda}
 \end{equation}
with  
 \begin{eqnarray}
&& \omega = \cB_N(1-\alpha)(1-\delta)\beta'\gamma'+\cC_N(1-\beta)(1-\gamma)\alpha'\delta' 
\nonumber\\
&&\xi = 
\cB_N\cC_N(1-\alpha)(1-\beta)(1-\gamma)(1-\delta)\frac{\alpha'\beta'\gamma'\delta'}{\widetilde{\alpha}\widetilde{\beta}\widetilde{\gamma}\widetilde{\delta}}
\nonumber\\
&&\eta = (\alpha\beta-\widetilde\alpha\widetilde\beta)(\gamma\delta-\widetilde\gamma\widetilde\delta)\xi \nonumber
\end{eqnarray}
and where
 \begin{eqnarray}
  \cA_N &=& \prod_{i=i}^{N-1} (f_{00}^{(i)}g_{00}^{(i)}), \\
  \cB_N &=& \prod_{i=1}^{N/2} \frac{f_{01}^{(2i-1)}f_{10}^{(2i-1)}g_{11}^{(2i-1)}}{\big(f_{00}^{(2i-1)}\big)^2 g_{00}^{(2i-1)}}
  \prod_{i=1}^{N/2-1} \frac{g_{01}^{(2i)}g_{10}^{(2i)}f_{11}^{(2i)}}{\big(g_{00}^{(2i)}\big)^2 f_{00}^{(2i)}}, \nonumber\\
  \cC_N &=& \prod_{i=1}^{N/2} \frac{g_{01}^{(2i-1)}g_{10}^{(2i-1)}f_{11}^{(2i-1)}}{\big(g_{00}^{(2i-1)}\big)^2 f_{00}^{(2i-1)}}
  \prod_{i=1}^{N/2-1} \frac{f_{01}^{(2i)}f_{10}^{(2i)}g_{11}^{(2j)}}{\big(f_{00}^{(2i)}\big)^2 g_{00}^{(2i)}}, \nonumber
 \end{eqnarray}
 and, $\alpha'=\alpha+\widetilde{\alpha}$, $\beta'=\beta+\widetilde{\beta}$, $\gamma'=\gamma+\widetilde{\gamma}$, $\delta'=\delta+\widetilde{\delta}$,
 \begin{equation*}
 \begin{aligned}
 & \widetilde\alpha = \frac{f_{10}^{(1)}g_{11}^{(1)}}{f_{00}^{(1)}g_{01}^{(1)}}(1-\alpha), \quad 
  \widetilde\beta = \frac{f_{11}^{(1)}g_{10}^{(1)}}{f_{01}^{(1)}g_{00}^{(1)}}(1-\beta), \\
 & \widetilde\gamma = \frac{f_{11}^{(N-1)}g_{01}^{(N-1)}}{f_{10}^{(N-1)}g_{00}^{(N-1)}}(1-\gamma), \quad 
  \widetilde\delta = \frac{f_{01}^{(N-1)}g_{11}^{(N-1)}}{f_{00}^{(N-1)}g_{10}^{(N-1)}}(1-\delta).
  \end{aligned}
  \end{equation*}
 When $s=0$, that is, in the non-deformed case, the polynomial factorizes as in \cite{Prosen2017} and the largest eigenvalue becomes $\lambda=1$
 as expected.

\begin{figure}[t]
\vspace{0mm}
\begin{center}
\includegraphics[width=\columnwidth]{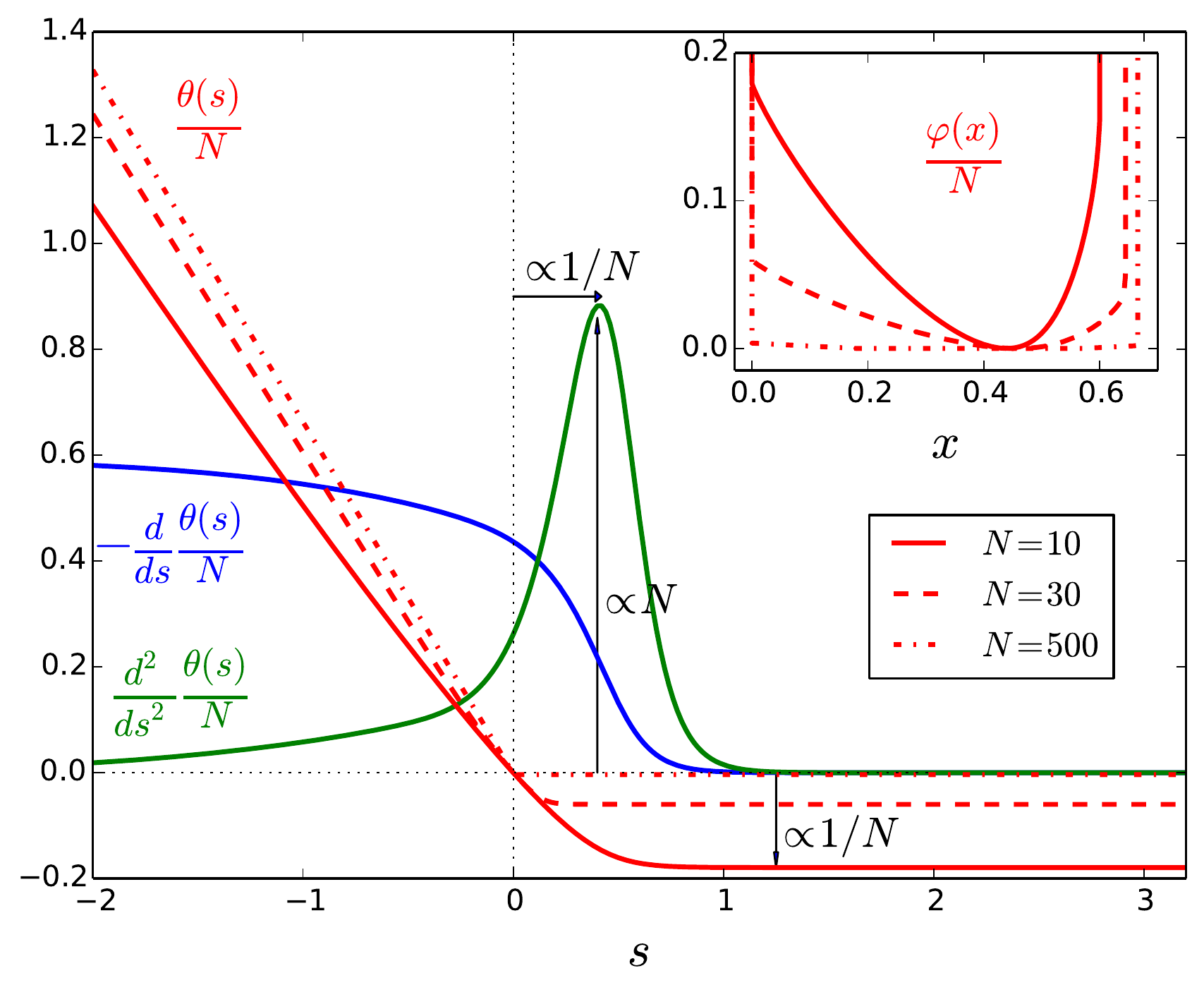}
\end{center}
\vspace{-5mm}
\caption{Dynamical phase transition in the RCA54:
Red curves show the exact SCGF $\theta(s)$ 
for the time-integrated number of up sites [$f_{j}(n,n')=\frac12(n+n')$ and $g_{j}=0$ in \er{obs}] for sizes $N=10,30,500$; the SCGF approaches the singular form \eqref{theta} for large size. The order parameter, $\lim_{T \to \infty} \langle \cO_T e^{-s \cO_T} \rangle / T N Z_T(s) = -\theta'(s)/N$ (blue) displays a first-order change between $s<0$ and $s>0$, while its susceptibility $\theta''(s)/N$ (green) diverges as $N$ (blue and green curves are for $N=10$). The first-order singularity at $s_c=0$ is approached as $1/N$. Inset: Exact rate function $\varphi(x)$ with $x=\cO_T / TN$ for sizes $N=10,30,500$. [Parameters are $(\alpha,\beta,\gamma,\delta)=(1/3,1/8,1/2,2/5)$.]
}
\label{fig:cumulants}
\end{figure}

\smallskip

\noindent
{\bf \em Dynamical phase transition.--}
From \er{lambda} we can obtain the behavior of the SCGF $\theta_N(s)$ in the large size limit. Since the observables we consider are extensive in system size, cf.\ \er{obs}, we have that $a:=-\lim_{N \to \infty}(\ln \cA_N)/(Ns)$, $b:=-\lim_{N \to \infty}(\ln \cB_N)/(Ns)$ and $c:=-\lim_{N \to \infty}(\ln \cC_N)/(Ns)$ exist
and are finite. The SCGF then takes the scaling form
\begin{equation} \label{eq:scaling}
 \theta_N(s) = \cE(N s)+\cO\Big(\frac{1}{N}\Big),
\end{equation}
where the function $\cE(\sigma)$ is defined such that $\exp[\cE(\sigma)]$ is the largest real root of the polynomial 
\begin{eqnarray} \label{eq:poly_scaling}
 & 0 & =  e^{4[\cE(\sigma)+\sigma a]}-\alpha\gamma e^{3[\cE(\sigma)+\sigma a]} \\
 &-& \Big[e^{-\sigma b} (1-\alpha)(1-\delta)+e^{-\sigma c}(1-\beta)(1-\gamma)\Big]e^{2[\cE(\sigma)+\sigma a]} \nonumber \\
  &-&e^{-(b+c)\sigma}\beta\delta e^{\cE(\sigma)+\sigma a} 
  + e^{-(b+c)\sigma}(\alpha+\beta-1)(\gamma+\delta-1). \nonumber 
 \end{eqnarray}
 The scaling form \eqref{eq:scaling} provides us immediately with system size behavior of the long-time cumulants of $\cO_T$
 \begin{equation}
 \lim_{T \to \infty} \frac{1}{T} \langle\!\langle \cO_T^k \rangle\!\rangle = \left. (-)^k \frac{d^k}{ds^k} \theta_N \right|_{s=0} \propto N^{k} 
 \end{equation}
where $\langle\!\langle \cdot \rangle\!\rangle$ indicates the cumulant. The supra-linear dependence on size for $k \geq 2$ indicates the presence of a singularity at $s=0$ in the large size limit. 

We can extract explicitly from \eqref{eq:poly_scaling} the exact asymptotic of the first few cumulants. From $k=1$ we get the average observable per unit time 
  \begin{equation}
\lim_{T \to \infty} \frac{1}{TN} \langle \cO_T \rangle =
 a+\frac{\mu b+ \nu c}{2(\mu+\nu)+\alpha\gamma-\beta \delta}+ \mathcal{O}\Big(\frac{1}{N}\Big)
 \end{equation}
while from $k=2$ the corresponding susceptibility 
\begin{align}
&\lim_{T \to \infty} \frac{1}{TN} {\rm var}~\cO_T  
= N \left[ 
-\frac{2bc(1-\alpha\gamma)+\mu b^2 +\nu c^2}{2(\mu+\nu)+\alpha\gamma-\beta \delta} 
\right.
\nonumber \\
 & +\frac{3(\mu b + \nu c)^2}{(2(\mu+\nu)+\alpha\gamma-\beta \delta)^2} 
 + \frac{2(b+c)(\mu b + \nu c)(2-\alpha\gamma)}{(2(\mu+\nu)+\alpha\gamma-\beta \delta)^2} 
 \nonumber \\
 &
 \left. 
 -\frac{2(\mu b + \nu c)^2(4+\mu+\nu-\alpha\gamma)}{(2(\mu+\nu)+\alpha\gamma-\beta \delta)^3}
 \right]
 + \mathcal{O}(1)
 \end{align}
 with $\mu = \gamma(1-\alpha) + \beta(1-\gamma)$ and $\nu = \delta(1-\alpha) + \alpha(1-\gamma)$.
  
The scaling function $\cE(\sigma)$ has the following properties: (i) at $\sigma=0$ it vanishes as the polynomial \eqref{eq:poly_scaling} trivially factorizes; (ii) it is a convex function and $\cE''(\sigma)$ admits a global maximum $\sigma^*$; (iii) if $(b+c)>0$ it has the asymptotic behavior
\begin{equation}
\cE(\sigma) = 
 \begin{cases}
  -a\sigma + \ln(\alpha\gamma) + o(1), \quad \sigma \to \infty \\
  -\frac{1}{3}(b+c+3a)\sigma + \frac{1}{3}\ln(\beta\delta) + o(1), \quad \sigma \to -\infty
 \end{cases}
 \nonumber
\end{equation}
[if $(b+c)<0$ the asymptotic behavior is obtained by $\sigma \to -\sigma$]. We can thus deduce that the SCGF converges to the limit shape
\begin{equation}
\lim_{N\to\infty}\frac{1}{N}\theta_N(s) = 
 \begin{cases}
  -a s , \quad s>0 \\
  -\frac{1}{3}(b+c+3a)s, \quad s<0
 \end{cases}
 \label{theta}
\end{equation}
when $(b+c)>0$ [for $(b+c)<0$ the shape is obtained by $s \to -s$]. The singularity at $s=0$ corresponds to a first-order phase transition.

Figure \ref{fig:cumulants} shows the SCGF for one choice of the observable (the time-integrated number of up sites). As $N$ grows, $\theta_N(s)$ approaches the piecewise linear form \eqref{theta}. The order parameter, $-\theta'(s)/N$, changes from a large value for $s$ negative to one close to zero for $s$ positive, the change becoming discontinuous for $N \to \infty$. The increasing sharpness of the crossover is manifested in the behaviour of the susceptibility $\theta''(s)/N$, whose peak grows as $N$. The value of $s$ at its peak indicates the location of the finite size crossover, which goes as $s_c \propto N^{-1}$. The finite size scaling of the transition point is similar to that expected in the FA model \cite{Bodineau2012,Bodineau2012b,Nemoto2017}, while the scaling of the susceptibility is different.  

The Inset to Fig.\ \ref{fig:cumulants} shows the rate function $\varphi(x)$ where $x=\cO_T / TN$ for various sizes, as obtained from the SCGF via the Legendre transform. As the size increases $\varphi(x)$ progressively broadens. For finite $N$ the broadening is indicative of large fluctuations, and a precursor of the phase transition. In the limit $N \to \infty$, it takes the shape of a flat square well, corresponding to the Maxwell construction due to the first-order coexistence of the two dynamical phases, the inactive one with $x_{\rm min} = a$ and the active one with $x_{\rm max} = \frac{1}{3}(b+c+3a)$, cf.\ \er{theta} ($\varphi=\infty$ elsewhere). Due to the integrable nature of the RCA54 -- and in contrast to facilitated models \cite{Garrahan2007,Garrahan2009,Garrahan2018} -- there are no fluctuations within each dynamical phase, which means that for finite $N$ the rate function should have the shape of a ``tilted ellipse'' \cite{Jordan2004,Lambert2015,Brandner2017}.

\begin{figure}[t]
\vspace{3mm}
\begin{center}
\includegraphics[width=\columnwidth]{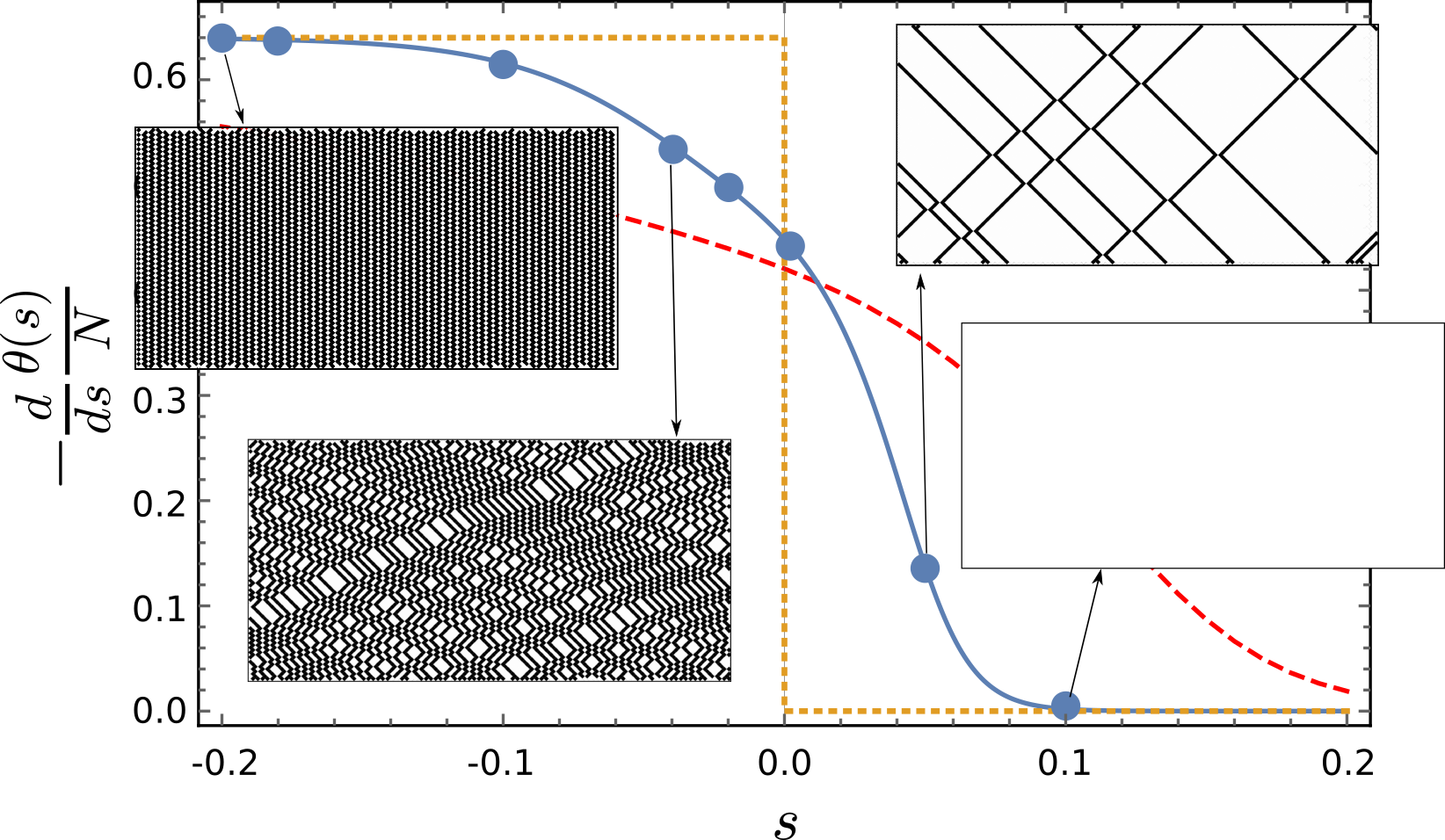}
\end{center}
\vspace{-7mm}
\caption{
Sampling of titled dynamics via Doob transform: The average under \er{eq:Doob} of the observable, $\langle \cO_T \rangle_{M_{\rm Doob}(s)} / T N$ (blue symbols), coincides with the exact value of the order parameter, $-\theta'(s)/N$ (blue curve), as shown for $N=100$. We show  sample trajectories for various values of $s$: For $s<0$, trajectories are dense in up sites, thus increasing activity; the leftmost trajectory maximises activity by becoming ordered is space and time - this is the arrangement in the inactive phase for $s<0$ in the $N \to \infty$ limit (cf.\ orange/dotted curve). For $s>0$ trajectories are sparse in up sites, thus reducing activity. 
(Same observable and parameters as in Fig.\ 2.)
}
\label{fig:active_inactive}
\end{figure}

\smallskip

\noindent
{\bf \em Doob transformation and optimal dynamics.--}
The dynamical phase transition above corresponds to a singular change at the level of fluctuations: if the ensemble of trajectories is reweighed by $e^{-s \cO_T}$ (the so-called $s$-ensemble \cite{Hedges2009,Garrahan2018}), there is a singular change in the nature of atypically active trajectories ($s<0$) to atypically inactive ones ($s>0$). These reweighed ensembles can be sampled from the original dynamics by post-processing, but this is exponentially costly in $T$. However, they can be optimally accessed in terms of an ``auxiliary'' \cite{Jack2010} or ``driven'' \cite{Chetrite2015} Markov process, by means of a so-called {\em generalised Doob transformation}; see also \cite{Borkar2003,Chetrite2015b,Jack2015b,Garrahan2016,Carollo2018,Derrida2018}.

From the matrix product construction of the leading eigenvector of $M(s)$ we can obtain the exact long-time Doob operator \cite{Jack2010,Chetrite2015,Garrahan2016}: 
\begin{equation} \label{eq:Doob}
M_{\rm Doob}=\frac{1}{\lambda(s)}\LL M(s) \LL^{-1} ,
\end{equation}
where $\LL$ is a diagonal operator formed out of components of 
the leading \emph{left} eigenvector $\bm{q}$ of $M(s)$, \textit{i.e.} $\LL_{\bf{n},\bf{n'}}:=\delta_{\bf{n},\bf{n'}} q_{\bf{n}}$.
The exact matrix product construction of $\bm{q}$ in given in \cite{SM}. The operator \eqref{eq:Doob} is a Markov matrix for a stochastic dynamics whose trajectories are guaranteed to coincide - for long-times - with those of the $s$-ensemble of the original $M$ \cite{Jack2010,Chetrite2015,Chetrite2015b,Jack2015b,Garrahan2016}. Figure \ref{fig:active_inactive} shows how $M_{\rm Doob}$ allows to sample the fluctuations of the RCA54 parameterised by $s$ in an optimal way.  

Note that due to the deterministic nature of the bulk dynamics of the RCA54, the Doob operator amounts to a non-trivial modification only of the boundary probabilities, which in $M_{\rm Doob}$ depend on the configuration of the whole lattice. For details see \cite{SM}.

\smallskip

\noindent
{\bf \em Conclusion.--} 
We studied the statistics of a general class of dynamical observables in a cellular automaton with stochastic boundary driving. We provided an exact expression of their scaled cumulant generating functions by means of an inhomogeneous matrix product expression for the leading eigenvector of the corresponding tilted Markov operator. Our results give a precise analytical description of the phase transition between active and inactive dynamical phases observed in a wide range of other models. 

We foresee extensions of our work here in several directions, including computing the large deviation statistics of currents, and even the complete ``level 2.5'' statistics for the empirical measure and fluxes \cite{Maes2008,Bertini2015b,Barato2015b,Hoppenau2016,Bertini2018}. The analytic inhomogeneous Matrix Ansatz introduced here could also be used to address similar questions in more complicated models, for instance for cellular automata with asymmetric constraints, and for systems with stochastic dynamics in the bulk.

\smallskip

\noindent
{\bf \em Acknowledgements.--} The work here has been supported by Advanced Grant 694544 -- OMNES of the European Research Council (ERC), Grants No.\ N1-0025, N1--0055 of the Slovenian Research Agency (ARRS), EPSRC programme grant EP/P009565/1, EPSRC Grant No.\ EP/R04421X/1, Leverhulme Grant RPG-2018-181.

\bibliographystyle{apsrev4-1}
\bibliography{activity-Bobenko}

\begin{thebibliography}{75}%
\makeatletter
\providecommand \@ifxundefined [1]{%
 \@ifx{#1\undefined}
}%
\providecommand \@ifnum [1]{%
 \ifnum #1\expandafter \@firstoftwo
 \else \expandafter \@secondoftwo
 \fi
}%
\providecommand \@ifx [1]{%
 \ifx #1\expandafter \@firstoftwo
 \else \expandafter \@secondoftwo
 \fi
}%
\providecommand \natexlab [1]{#1}%
\providecommand \enquote  [1]{``#1''}%
\providecommand \bibnamefont  [1]{#1}%
\providecommand \bibfnamefont [1]{#1}%
\providecommand \citenamefont [1]{#1}%
\providecommand \href@noop [0]{\@secondoftwo}%
\providecommand \href [0]{\begingroup \@sanitize@url \@href}%
\providecommand \@href[1]{\@@startlink{#1}\@@href}%
\providecommand \@@href[1]{\endgroup#1\@@endlink}%
\providecommand \@sanitize@url [0]{\catcode `\\12\catcode `\$12\catcode
  `\&12\catcode `\#12\catcode `\^12\catcode `\_12\catcode `\%12\relax}%
\providecommand \@@startlink[1]{}%
\providecommand \@@endlink[0]{}%
\providecommand \url  [0]{\begingroup\@sanitize@url \@url }%
\providecommand \@url [1]{\endgroup\@href {#1}{\urlprefix }}%
\providecommand \urlprefix  [0]{URL }%
\providecommand \Eprint [0]{\href }%
\providecommand \doibase [0]{http://dx.doi.org/}%
\providecommand \selectlanguage [0]{\@gobble}%
\providecommand \bibinfo  [0]{\@secondoftwo}%
\providecommand \bibfield  [0]{\@secondoftwo}%
\providecommand \translation [1]{[#1]}%
\providecommand \BibitemOpen [0]{}%
\providecommand \bibitemStop [0]{}%
\providecommand \bibitemNoStop [0]{.\EOS\space}%
\providecommand \EOS [0]{\spacefactor3000\relax}%
\providecommand \BibitemShut  [1]{\csname bibitem#1\endcsname}%
\let\auto@bib@innerbib\@empty
\bibitem [{\citenamefont {Derrida}(2007)}]{Derrida2007}%
  \BibitemOpen
  \bibfield  {author} {\bibinfo {author} {\bibfnamefont {B.}~\bibnamefont
  {Derrida}},\ }\href {http://stacks.iop.org/1742-5468/2007/i=07/a=P07023}
  {\bibfield  {journal} {\bibinfo  {journal} {J. Stat. Mech.}\ }\textbf
  {\bibinfo {volume} {2007}},\ \bibinfo {pages} {P07023} (\bibinfo {year}
  {2007})}\BibitemShut {NoStop}%
\bibitem [{\citenamefont {Mallick}(2015)}]{Mallick2015}%
  \BibitemOpen
  \bibfield  {author} {\bibinfo {author} {\bibfnamefont {K.}~\bibnamefont
  {Mallick}},\ }\href {\doibase https://doi.org/10.1016/j.physa.2014.07.046}
  {\bibfield  {journal} {\bibinfo  {journal} {Physica A: Statistical Mechanics
  and its Applications}\ }\textbf {\bibinfo {volume} {418}},\ \bibinfo {pages}
  {17 } (\bibinfo {year} {2015})},\ \bibinfo {note} {proceedings of the 13th
  International Summer School on Fundamental Problems in Statistical
  Physics}\BibitemShut {NoStop}%
\bibitem [{\citenamefont {Henley}(2010)}]{Henley2010}%
  \BibitemOpen
  \bibfield  {author} {\bibinfo {author} {\bibfnamefont {C.~L.}\ \bibnamefont
  {Henley}},\ }\href@noop {} {\bibfield  {journal} {\bibinfo  {journal} {Annu.
  Rev. Condens. Matter Phys.}\ }\textbf {\bibinfo {volume} {1}},\ \bibinfo
  {pages} {179} (\bibinfo {year} {2010})}\BibitemShut {NoStop}%
\bibitem [{\citenamefont {Chalker}(2017)}]{Chalker2017}%
  \BibitemOpen
  \bibfield  {author} {\bibinfo {author} {\bibfnamefont {J.~T.}\ \bibnamefont
  {Chalker}},\ }\href@noop {} {\bibfield  {journal} {\bibinfo  {journal}
  {Topological Aspects of Condensed Matter Physics: Lecture Notes of the Les
  Houches Summer School: Volume 103, August 2014}\ }\textbf {\bibinfo {volume}
  {103}},\ \bibinfo {pages} {123} (\bibinfo {year} {2017})}\BibitemShut
  {NoStop}%
\bibitem [{\citenamefont {Ritort}\ and\ \citenamefont
  {Sollich}(2003)}]{Ritort2003}%
  \BibitemOpen
  \bibfield  {author} {\bibinfo {author} {\bibfnamefont {F.}~\bibnamefont
  {Ritort}}\ and\ \bibinfo {author} {\bibfnamefont {P.}~\bibnamefont
  {Sollich}},\ }\href@noop {} {\bibfield  {journal} {\bibinfo  {journal} {Adv.
  Phys.}\ }\textbf {\bibinfo {volume} {52}},\ \bibinfo {pages} {219} (\bibinfo
  {year} {2003})}\BibitemShut {NoStop}%
\bibitem [{\citenamefont {Garrahan}\ \emph {et~al.}(2011)\citenamefont
  {Garrahan}, \citenamefont {Sollich},\ and\ \citenamefont
  {Toninelli}}]{Garrahan2011}%
  \BibitemOpen
  \bibfield  {author} {\bibinfo {author} {\bibfnamefont {J.~P.}\ \bibnamefont
  {Garrahan}}, \bibinfo {author} {\bibfnamefont {P.}~\bibnamefont {Sollich}}, \
  and\ \bibinfo {author} {\bibfnamefont {C.}~\bibnamefont {Toninelli}},\ }in\
  \href@noop {} {\emph {\bibinfo {booktitle} {Dynamical Heterogeneities in
  Glasses, Colloids, and Granular Media}}},\ \bibinfo {series and number}
  {International Series of Monographs on Physics},\ \bibinfo {editor} {edited
  by\ \bibinfo {editor} {\bibfnamefont {L.}~\bibnamefont {Berthier}}, \bibinfo
  {editor} {\bibfnamefont {G.}~\bibnamefont {Biroli}}, \bibinfo {editor}
  {\bibfnamefont {J.-P.}\ \bibnamefont {Bouchaud}}, \bibinfo {editor}
  {\bibfnamefont {L.}~\bibnamefont {Cipelletti}}, \ and\ \bibinfo {editor}
  {\bibfnamefont {W.}~\bibnamefont {van Saarloos}}}\ (\bibinfo  {publisher}
  {Oxford University Press},\ \bibinfo {address} {Oxford, UK},\ \bibinfo {year}
  {2011})\BibitemShut {NoStop}%
\bibitem [{\citenamefont {van Horssen}\ \emph {et~al.}(2015)\citenamefont {van
  Horssen}, \citenamefont {Levi},\ and\ \citenamefont
  {Garrahan}}]{Horssen2015}%
  \BibitemOpen
  \bibfield  {author} {\bibinfo {author} {\bibfnamefont {M.}~\bibnamefont {van
  Horssen}}, \bibinfo {author} {\bibfnamefont {E.}~\bibnamefont {Levi}}, \ and\
  \bibinfo {author} {\bibfnamefont {J.~P.}\ \bibnamefont {Garrahan}},\ }\href
  {\doibase 10.1103/PhysRevB.92.100305} {\bibfield  {journal} {\bibinfo
  {journal} {Phys. Rev. B}\ }\textbf {\bibinfo {volume} {92}},\ \bibinfo
  {pages} {100305} (\bibinfo {year} {2015})}\BibitemShut {NoStop}%
\bibitem [{\citenamefont {Smith}\ \emph {et~al.}(2017)\citenamefont {Smith},
  \citenamefont {Knolle}, \citenamefont {Kovrizhin},\ and\ \citenamefont
  {Moessner}}]{Smith2017}%
  \BibitemOpen
  \bibfield  {author} {\bibinfo {author} {\bibfnamefont {A.}~\bibnamefont
  {Smith}}, \bibinfo {author} {\bibfnamefont {J.}~\bibnamefont {Knolle}},
  \bibinfo {author} {\bibfnamefont {D.~L.}\ \bibnamefont {Kovrizhin}}, \ and\
  \bibinfo {author} {\bibfnamefont {R.}~\bibnamefont {Moessner}},\ }\href@noop
  {} {\bibfield  {journal} {\bibinfo  {journal} {Phys. Rev. Lett.}\ }\textbf
  {\bibinfo {volume} {118}},\ \bibinfo {pages} {266601} (\bibinfo {year}
  {2017})}\BibitemShut {NoStop}%
\bibitem [{\citenamefont {Shiraishi}\ and\ \citenamefont
  {Mori}(2017)}]{Shiraishi2017}%
  \BibitemOpen
  \bibfield  {author} {\bibinfo {author} {\bibfnamefont {N.}~\bibnamefont
  {Shiraishi}}\ and\ \bibinfo {author} {\bibfnamefont {T.}~\bibnamefont
  {Mori}},\ }\href {\doibase 10.1103/PhysRevLett.119.030601} {\bibfield
  {journal} {\bibinfo  {journal} {Phys. Rev. Lett.}\ }\textbf {\bibinfo
  {volume} {119}},\ \bibinfo {pages} {030601} (\bibinfo {year}
  {2017})}\BibitemShut {NoStop}%
\bibitem [{\citenamefont {Lan}\ \emph {et~al.}(2018)\citenamefont {Lan},
  \citenamefont {van Horssen}, \citenamefont {Powell},\ and\ \citenamefont
  {Garrahan}}]{Lan2018}%
  \BibitemOpen
  \bibfield  {author} {\bibinfo {author} {\bibfnamefont {Z.}~\bibnamefont
  {Lan}}, \bibinfo {author} {\bibfnamefont {M.}~\bibnamefont {van Horssen}},
  \bibinfo {author} {\bibfnamefont {S.}~\bibnamefont {Powell}}, \ and\ \bibinfo
  {author} {\bibfnamefont {J.~P.}\ \bibnamefont {Garrahan}},\ }\href {\doibase
  10.1103/PhysRevLett.121.040603} {\bibfield  {journal} {\bibinfo  {journal}
  {Phys. Rev. Lett.}\ }\textbf {\bibinfo {volume} {121}},\ \bibinfo {pages}
  {040603} (\bibinfo {year} {2018})}\BibitemShut {NoStop}%
\bibitem [{\citenamefont {Turner}\ \emph {et~al.}(2018)\citenamefont {Turner},
  \citenamefont {Michailidis}, \citenamefont {Abanin}, \citenamefont {Serbyn},\
  and\ \citenamefont {Papi{\'c}}}]{Turner2018}%
  \BibitemOpen
  \bibfield  {author} {\bibinfo {author} {\bibfnamefont {C.}~\bibnamefont
  {Turner}}, \bibinfo {author} {\bibfnamefont {A.}~\bibnamefont {Michailidis}},
  \bibinfo {author} {\bibfnamefont {D.}~\bibnamefont {Abanin}}, \bibinfo
  {author} {\bibfnamefont {M.}~\bibnamefont {Serbyn}}, \ and\ \bibinfo {author}
  {\bibfnamefont {Z.}~\bibnamefont {Papi{\'c}}},\ }\href@noop {} {\bibfield
  {journal} {\bibinfo  {journal} {Nature Physics}\ }\textbf {\bibinfo {volume}
  {14}},\ \bibinfo {pages} {745} (\bibinfo {year} {2018})}\BibitemShut
  {NoStop}%
\bibitem [{\citenamefont {Nahum}\ \emph {et~al.}(2017)\citenamefont {Nahum},
  \citenamefont {Ruhman}, \citenamefont {Vijay},\ and\ \citenamefont
  {Haah}}]{Nahum2017}%
  \BibitemOpen
  \bibfield  {author} {\bibinfo {author} {\bibfnamefont {A.}~\bibnamefont
  {Nahum}}, \bibinfo {author} {\bibfnamefont {J.}~\bibnamefont {Ruhman}},
  \bibinfo {author} {\bibfnamefont {S.}~\bibnamefont {Vijay}}, \ and\ \bibinfo
  {author} {\bibfnamefont {J.}~\bibnamefont {Haah}},\ }\href@noop {} {\bibfield
   {journal} {\bibinfo  {journal} {Phys. Rev. X}\ }\textbf {\bibinfo {volume}
  {7}},\ \bibinfo {pages} {031016} (\bibinfo {year} {2017})}\BibitemShut
  {NoStop}%
\bibitem [{\citenamefont {von Keyserlingk}\ \emph {et~al.}(2018)\citenamefont
  {von Keyserlingk}, \citenamefont {Rakovszky}, \citenamefont {Pollmann},\ and\
  \citenamefont {Sondhi}}]{Keyserlingk2018}%
  \BibitemOpen
  \bibfield  {author} {\bibinfo {author} {\bibfnamefont {C.~W.}\ \bibnamefont
  {von Keyserlingk}}, \bibinfo {author} {\bibfnamefont {T.}~\bibnamefont
  {Rakovszky}}, \bibinfo {author} {\bibfnamefont {F.}~\bibnamefont {Pollmann}},
  \ and\ \bibinfo {author} {\bibfnamefont {S.~L.}\ \bibnamefont {Sondhi}},\
  }\href@noop {} {\bibfield  {journal} {\bibinfo  {journal} {Phys. Rev. X}\
  }\textbf {\bibinfo {volume} {8}},\ \bibinfo {pages} {021013} (\bibinfo {year}
  {2018})}\BibitemShut {NoStop}%
\bibitem [{\citenamefont {Rowlands}\ and\ \citenamefont
  {Lamacraft}(2018)}]{Rowlands2018}%
  \BibitemOpen
  \bibfield  {author} {\bibinfo {author} {\bibfnamefont {D.~A.}\ \bibnamefont
  {Rowlands}}\ and\ \bibinfo {author} {\bibfnamefont {A.}~\bibnamefont
  {Lamacraft}},\ }\href {\doibase 10.1103/PhysRevB.98.195125} {\bibfield
  {journal} {\bibinfo  {journal} {Phys. Rev. B}\ }\textbf {\bibinfo {volume}
  {98}},\ \bibinfo {pages} {195125} (\bibinfo {year} {2018})}\BibitemShut
  {NoStop}%
\bibitem [{\citenamefont {Chen}\ and\ \citenamefont {Zhou}(2018)}]{Chen2018}%
  \BibitemOpen
  \bibfield  {author} {\bibinfo {author} {\bibfnamefont {X.}~\bibnamefont
  {Chen}}\ and\ \bibinfo {author} {\bibfnamefont {T.}~\bibnamefont {Zhou}},\
  }\href@noop {} {\bibfield  {journal} {\bibinfo  {journal} {arXiv:1808.09812}\
  } (\bibinfo {year} {2018})}\BibitemShut {NoStop}%
\bibitem [{\citenamefont {Gopalakrishnan}(2018)}]{Gopalakrishnan2018}%
  \BibitemOpen
  \bibfield  {author} {\bibinfo {author} {\bibfnamefont {S.}~\bibnamefont
  {Gopalakrishnan}},\ }\href {\doibase 10.1103/PhysRevB.98.060302} {\bibfield
  {journal} {\bibinfo  {journal} {Phys. Rev. B}\ }\textbf {\bibinfo {volume}
  {98}},\ \bibinfo {pages} {060302} (\bibinfo {year} {2018})}\BibitemShut
  {NoStop}%
\bibitem [{\citenamefont {Knap}(2018)}]{Knap2018}%
  \BibitemOpen
  \bibfield  {author} {\bibinfo {author} {\bibfnamefont {M.}~\bibnamefont
  {Knap}},\ }\href {\doibase 10.1103/PhysRevB.98.184416} {\bibfield  {journal}
  {\bibinfo  {journal} {Phys. Rev. B}\ }\textbf {\bibinfo {volume} {98}},\
  \bibinfo {pages} {184416} (\bibinfo {year} {2018})}\BibitemShut {NoStop}%
\bibitem [{\citenamefont {Tran}\ \emph {et~al.}(2018)\citenamefont {Tran},
  \citenamefont {Guo}, \citenamefont {Su}, \citenamefont {Garrison},
  \citenamefont {Eldredge}, \citenamefont {Foss-Feig}, \citenamefont {Childs},\
  and\ \citenamefont {Gorshkov}}]{Tran2018}%
  \BibitemOpen
  \bibfield  {author} {\bibinfo {author} {\bibfnamefont {M.~C.}\ \bibnamefont
  {Tran}}, \bibinfo {author} {\bibfnamefont {A.~Y.}\ \bibnamefont {Guo}},
  \bibinfo {author} {\bibfnamefont {Y.}~\bibnamefont {Su}}, \bibinfo {author}
  {\bibfnamefont {J.~R.}\ \bibnamefont {Garrison}}, \bibinfo {author}
  {\bibfnamefont {Z.}~\bibnamefont {Eldredge}}, \bibinfo {author}
  {\bibfnamefont {M.}~\bibnamefont {Foss-Feig}}, \bibinfo {author}
  {\bibfnamefont {A.~M.}\ \bibnamefont {Childs}}, \ and\ \bibinfo {author}
  {\bibfnamefont {A.~V.}\ \bibnamefont {Gorshkov}},\ }\href@noop {} {\bibfield
  {journal} {\bibinfo  {journal} {arXiv:1808.05225}\ } (\bibinfo {year}
  {2018})}\BibitemShut {NoStop}%
\bibitem [{\citenamefont {Gopalakrishnan}\ \emph {et~al.}(2018)\citenamefont
  {Gopalakrishnan}, \citenamefont {Huse}, \citenamefont {Khemani},\ and\
  \citenamefont {Vasseur}}]{Gopalakrishnan2018b}%
  \BibitemOpen
  \bibfield  {author} {\bibinfo {author} {\bibfnamefont {S.}~\bibnamefont
  {Gopalakrishnan}}, \bibinfo {author} {\bibfnamefont {D.~A.}\ \bibnamefont
  {Huse}}, \bibinfo {author} {\bibfnamefont {V.}~\bibnamefont {Khemani}}, \
  and\ \bibinfo {author} {\bibfnamefont {R.}~\bibnamefont {Vasseur}},\ }\href
  {\doibase 10.1103/PhysRevB.98.220303} {\bibfield  {journal} {\bibinfo
  {journal} {Phys. Rev. B}\ }\textbf {\bibinfo {volume} {98}},\ \bibinfo
  {pages} {220303} (\bibinfo {year} {2018})}\BibitemShut {NoStop}%
\bibitem [{\citenamefont {Lesanovsky}\ and\ \citenamefont
  {Garrahan}(2013)}]{Lesanovsky2013}%
  \BibitemOpen
  \bibfield  {author} {\bibinfo {author} {\bibfnamefont {I.}~\bibnamefont
  {Lesanovsky}}\ and\ \bibinfo {author} {\bibfnamefont {J.~P.}\ \bibnamefont
  {Garrahan}},\ }\href {\doibase 10.1103/PhysRevLett.111.215305} {\bibfield
  {journal} {\bibinfo  {journal} {Phys. Rev. Lett.}\ }\textbf {\bibinfo
  {volume} {111}},\ \bibinfo {pages} {215305} (\bibinfo {year}
  {2013})}\BibitemShut {NoStop}%
\bibitem [{\citenamefont {Urvoy}\ \emph {et~al.}(2015)\citenamefont {Urvoy},
  \citenamefont {Ripka}, \citenamefont {Lesanovsky}, \citenamefont {Booth},
  \citenamefont {Shaffer}, \citenamefont {Pfau},\ and\ \citenamefont
  {L\"ow}}]{Urvoy2015}%
  \BibitemOpen
  \bibfield  {author} {\bibinfo {author} {\bibfnamefont {A.}~\bibnamefont
  {Urvoy}}, \bibinfo {author} {\bibfnamefont {F.}~\bibnamefont {Ripka}},
  \bibinfo {author} {\bibfnamefont {I.}~\bibnamefont {Lesanovsky}}, \bibinfo
  {author} {\bibfnamefont {D.}~\bibnamefont {Booth}}, \bibinfo {author}
  {\bibfnamefont {J.~P.}\ \bibnamefont {Shaffer}}, \bibinfo {author}
  {\bibfnamefont {T.}~\bibnamefont {Pfau}}, \ and\ \bibinfo {author}
  {\bibfnamefont {R.}~\bibnamefont {L\"ow}},\ }\href {\doibase
  10.1103/PhysRevLett.114.203002} {\bibfield  {journal} {\bibinfo  {journal}
  {Phys. Rev. Lett.}\ }\textbf {\bibinfo {volume} {114}},\ \bibinfo {pages}
  {203002} (\bibinfo {year} {2015})}\BibitemShut {NoStop}%
\bibitem [{\citenamefont {Valado}\ \emph {et~al.}(2016)\citenamefont {Valado},
  \citenamefont {Simonelli}, \citenamefont {Hoogerland}, \citenamefont
  {Lesanovsky}, \citenamefont {Garrahan}, \citenamefont {Arimondo},
  \citenamefont {Ciampini},\ and\ \citenamefont {Morsch}}]{Valado2016}%
  \BibitemOpen
  \bibfield  {author} {\bibinfo {author} {\bibfnamefont {M.~M.}\ \bibnamefont
  {Valado}}, \bibinfo {author} {\bibfnamefont {C.}~\bibnamefont {Simonelli}},
  \bibinfo {author} {\bibfnamefont {M.~D.}\ \bibnamefont {Hoogerland}},
  \bibinfo {author} {\bibfnamefont {I.}~\bibnamefont {Lesanovsky}}, \bibinfo
  {author} {\bibfnamefont {J.~P.}\ \bibnamefont {Garrahan}}, \bibinfo {author}
  {\bibfnamefont {E.}~\bibnamefont {Arimondo}}, \bibinfo {author}
  {\bibfnamefont {D.}~\bibnamefont {Ciampini}}, \ and\ \bibinfo {author}
  {\bibfnamefont {O.}~\bibnamefont {Morsch}},\ }\href {\doibase
  10.1103/PhysRevA.93.040701} {\bibfield  {journal} {\bibinfo  {journal} {Phys.
  Rev. A}\ }\textbf {\bibinfo {volume} {93}},\ \bibinfo {pages} {040701}
  (\bibinfo {year} {2016})}\BibitemShut {NoStop}%
\bibitem [{\citenamefont {Lecomte}\ \emph {et~al.}(2007)\citenamefont
  {Lecomte}, \citenamefont {Appert-Rolland},\ and\ \citenamefont {van
  Wijland}}]{Lecomte2007}%
  \BibitemOpen
  \bibfield  {author} {\bibinfo {author} {\bibfnamefont {V.}~\bibnamefont
  {Lecomte}}, \bibinfo {author} {\bibfnamefont {C.}~\bibnamefont
  {Appert-Rolland}}, \ and\ \bibinfo {author} {\bibfnamefont {F.}~\bibnamefont
  {van Wijland}},\ }\href@noop {} {\bibfield  {journal} {\bibinfo  {journal}
  {J. Stat. Phys.}\ }\textbf {\bibinfo {volume} {127}},\ \bibinfo {pages} {51}
  (\bibinfo {year} {2007})}\BibitemShut {NoStop}%
\bibitem [{\citenamefont {Garrahan}\ \emph {et~al.}(2007)\citenamefont
  {Garrahan}, \citenamefont {Jack}, \citenamefont {Lecomte}, \citenamefont
  {Pitard}, \citenamefont {van Duijvendijk},\ and\ \citenamefont {van
  Wijland}}]{Garrahan2007}%
  \BibitemOpen
  \bibfield  {author} {\bibinfo {author} {\bibfnamefont {J.~P.}\ \bibnamefont
  {Garrahan}}, \bibinfo {author} {\bibfnamefont {R.~L.}\ \bibnamefont {Jack}},
  \bibinfo {author} {\bibfnamefont {V.}~\bibnamefont {Lecomte}}, \bibinfo
  {author} {\bibfnamefont {E.}~\bibnamefont {Pitard}}, \bibinfo {author}
  {\bibfnamefont {K.}~\bibnamefont {van Duijvendijk}}, \ and\ \bibinfo {author}
  {\bibfnamefont {F.}~\bibnamefont {van Wijland}},\ }\href@noop {} {\bibfield
  {journal} {\bibinfo  {journal} {Phys. Rev. Lett.}\ }\textbf {\bibinfo
  {volume} {98}},\ \bibinfo {pages} {195702} (\bibinfo {year}
  {2007})}\BibitemShut {NoStop}%
\bibitem [{\citenamefont {Garrahan}\ \emph {et~al.}(2009)\citenamefont
  {Garrahan}, \citenamefont {Jack}, \citenamefont {Lecomte}, \citenamefont
  {Pitard}, \citenamefont {van Duijvendijk},\ and\ \citenamefont {van
  Wijland}}]{Garrahan2009}%
  \BibitemOpen
  \bibfield  {author} {\bibinfo {author} {\bibfnamefont {J.~P.}\ \bibnamefont
  {Garrahan}}, \bibinfo {author} {\bibfnamefont {R.~L.}\ \bibnamefont {Jack}},
  \bibinfo {author} {\bibfnamefont {V.}~\bibnamefont {Lecomte}}, \bibinfo
  {author} {\bibfnamefont {E.}~\bibnamefont {Pitard}}, \bibinfo {author}
  {\bibfnamefont {K.}~\bibnamefont {van Duijvendijk}}, \ and\ \bibinfo {author}
  {\bibfnamefont {F.}~\bibnamefont {van Wijland}},\ }\href@noop {} {\bibfield
  {journal} {\bibinfo  {journal} {J. Phys. A}\ }\textbf {\bibinfo {volume}
  {42}},\ \bibinfo {pages} {075007} (\bibinfo {year} {2009})}\BibitemShut
  {NoStop}%
\bibitem [{\citenamefont {Touchette}(2009)}]{Touchette2009}%
  \BibitemOpen
  \bibfield  {author} {\bibinfo {author} {\bibfnamefont {H.}~\bibnamefont
  {Touchette}},\ }\href@noop {} {\bibfield  {journal} {\bibinfo  {journal}
  {Phys. Rep.}\ }\textbf {\bibinfo {volume} {478}},\ \bibinfo {pages} {1}
  (\bibinfo {year} {2009})}\BibitemShut {NoStop}%
\bibitem [{\citenamefont {Appert-Rolland}\ \emph {et~al.}(2008)\citenamefont
  {Appert-Rolland}, \citenamefont {Derrida}, \citenamefont {Lecomte},\ and\
  \citenamefont {van Wijland}}]{Appert-Rolland2008}%
  \BibitemOpen
  \bibfield  {author} {\bibinfo {author} {\bibfnamefont {C.}~\bibnamefont
  {Appert-Rolland}}, \bibinfo {author} {\bibfnamefont {B.}~\bibnamefont
  {Derrida}}, \bibinfo {author} {\bibfnamefont {V.}~\bibnamefont {Lecomte}}, \
  and\ \bibinfo {author} {\bibfnamefont {F.}~\bibnamefont {van Wijland}},\
  }\href {\doibase 10.1103/PhysRevE.78.021122} {\bibfield  {journal} {\bibinfo
  {journal} {Phys. Rev. E}\ }\textbf {\bibinfo {volume} {78}},\ \bibinfo
  {pages} {021122} (\bibinfo {year} {2008})}\BibitemShut {NoStop}%
\bibitem [{\citenamefont {Espigares}\ \emph {et~al.}(2013)\citenamefont
  {Espigares}, \citenamefont {Garrido},\ and\ \citenamefont
  {Hurtado}}]{Espigares2013}%
  \BibitemOpen
  \bibfield  {author} {\bibinfo {author} {\bibfnamefont {C.~P.}\ \bibnamefont
  {Espigares}}, \bibinfo {author} {\bibfnamefont {P.~L.}\ \bibnamefont
  {Garrido}}, \ and\ \bibinfo {author} {\bibfnamefont {P.~I.}\ \bibnamefont
  {Hurtado}},\ }\href {\doibase 10.1103/PhysRevE.87.032115} {\bibfield
  {journal} {\bibinfo  {journal} {Phys. Rev. E}\ }\textbf {\bibinfo {volume}
  {87}},\ \bibinfo {pages} {032115} (\bibinfo {year} {2013})}\BibitemShut
  {NoStop}%
\bibitem [{\citenamefont {Jack}\ \emph {et~al.}(2015)\citenamefont {Jack},
  \citenamefont {Thompson},\ and\ \citenamefont {Sollich}}]{Jack2015}%
  \BibitemOpen
  \bibfield  {author} {\bibinfo {author} {\bibfnamefont {R.~L.}\ \bibnamefont
  {Jack}}, \bibinfo {author} {\bibfnamefont {I.~R.}\ \bibnamefont {Thompson}},
  \ and\ \bibinfo {author} {\bibfnamefont {P.}~\bibnamefont {Sollich}},\ }\href
  {\doibase 10.1103/PhysRevLett.114.060601} {\bibfield  {journal} {\bibinfo
  {journal} {Phys. Rev. Lett.}\ }\textbf {\bibinfo {volume} {114}},\ \bibinfo
  {pages} {060601} (\bibinfo {year} {2015})}\BibitemShut {NoStop}%
\bibitem [{\citenamefont {Karevski}\ and\ \citenamefont
  {Sch\"utz}(2017)}]{Karevski2017}%
  \BibitemOpen
  \bibfield  {author} {\bibinfo {author} {\bibfnamefont {D.}~\bibnamefont
  {Karevski}}\ and\ \bibinfo {author} {\bibfnamefont {G.~M.}\ \bibnamefont
  {Sch\"utz}},\ }\href@noop {} {\bibfield  {journal} {\bibinfo  {journal}
  {Phys. Rev. Lett.}\ }\textbf {\bibinfo {volume} {118}},\ \bibinfo {pages}
  {030601} (\bibinfo {year} {2017})}\BibitemShut {NoStop}%
\bibitem [{\citenamefont {Oakes}\ \emph {et~al.}(2018)\citenamefont {Oakes},
  \citenamefont {Powell}, \citenamefont {Castelnovo}, \citenamefont
  {Lamacraft},\ and\ \citenamefont {Garrahan}}]{Oakes2018}%
  \BibitemOpen
  \bibfield  {author} {\bibinfo {author} {\bibfnamefont {T.}~\bibnamefont
  {Oakes}}, \bibinfo {author} {\bibfnamefont {S.}~\bibnamefont {Powell}},
  \bibinfo {author} {\bibfnamefont {C.}~\bibnamefont {Castelnovo}}, \bibinfo
  {author} {\bibfnamefont {A.}~\bibnamefont {Lamacraft}}, \ and\ \bibinfo
  {author} {\bibfnamefont {J.~P.}\ \bibnamefont {Garrahan}},\ }\href {\doibase
  10.1103/PhysRevB.98.064302} {\bibfield  {journal} {\bibinfo  {journal} {Phys.
  Rev. B}\ }\textbf {\bibinfo {volume} {98}},\ \bibinfo {pages} {064302}
  (\bibinfo {year} {2018})}\BibitemShut {NoStop}%
\bibitem [{\citenamefont {Hedges}\ \emph {et~al.}(2009)\citenamefont {Hedges},
  \citenamefont {Jack}, \citenamefont {Garrahan},\ and\ \citenamefont
  {Chandler}}]{Hedges2009}%
  \BibitemOpen
  \bibfield  {author} {\bibinfo {author} {\bibfnamefont {L.~O.}\ \bibnamefont
  {Hedges}}, \bibinfo {author} {\bibfnamefont {R.~L.}\ \bibnamefont {Jack}},
  \bibinfo {author} {\bibfnamefont {J.~P.}\ \bibnamefont {Garrahan}}, \ and\
  \bibinfo {author} {\bibfnamefont {D.}~\bibnamefont {Chandler}},\ }\href@noop
  {} {\bibfield  {journal} {\bibinfo  {journal} {Science}\ }\textbf {\bibinfo
  {volume} {323}},\ \bibinfo {pages} {1309} (\bibinfo {year}
  {2009})}\BibitemShut {NoStop}%
\bibitem [{\citenamefont {Speck}\ \emph {et~al.}(2012)\citenamefont {Speck},
  \citenamefont {Malins},\ and\ \citenamefont {Royall}}]{Speck2012}%
  \BibitemOpen
  \bibfield  {author} {\bibinfo {author} {\bibfnamefont {T.}~\bibnamefont
  {Speck}}, \bibinfo {author} {\bibfnamefont {A.}~\bibnamefont {Malins}}, \
  and\ \bibinfo {author} {\bibfnamefont {C.~P.}\ \bibnamefont {Royall}},\
  }\href@noop {} {\bibfield  {journal} {\bibinfo  {journal} {Phys. Rev. Lett.}\
  }\textbf {\bibinfo {volume} {109}},\ \bibinfo {pages} {195703} (\bibinfo
  {year} {2012})}\BibitemShut {NoStop}%
\bibitem [{\citenamefont {Weber}\ \emph {et~al.}(2013)\citenamefont {Weber},
  \citenamefont {Jack},\ and\ \citenamefont {Pande}}]{Weber2013}%
  \BibitemOpen
  \bibfield  {author} {\bibinfo {author} {\bibfnamefont {J.~K.}\ \bibnamefont
  {Weber}}, \bibinfo {author} {\bibfnamefont {R.~L.}\ \bibnamefont {Jack}}, \
  and\ \bibinfo {author} {\bibfnamefont {V.~S.}\ \bibnamefont {Pande}},\
  }\href@noop {} {\bibfield  {journal} {\bibinfo  {journal} {J. Am. Chem.
  Soc.}\ }\textbf {\bibinfo {volume} {135}},\ \bibinfo {pages} {5501} (\bibinfo
  {year} {2013})}\BibitemShut {NoStop}%
\bibitem [{\citenamefont {Baek}\ \emph {et~al.}(2017)\citenamefont {Baek},
  \citenamefont {Kafri},\ and\ \citenamefont {Lecomte}}]{Baek2017}%
  \BibitemOpen
  \bibfield  {author} {\bibinfo {author} {\bibfnamefont {Y.}~\bibnamefont
  {Baek}}, \bibinfo {author} {\bibfnamefont {Y.}~\bibnamefont {Kafri}}, \ and\
  \bibinfo {author} {\bibfnamefont {V.}~\bibnamefont {Lecomte}},\ }\href
  {\doibase 10.1103/PhysRevLett.118.030604} {\bibfield  {journal} {\bibinfo
  {journal} {Phys. Rev. Lett.}\ }\textbf {\bibinfo {volume} {118}},\ \bibinfo
  {pages} {030604} (\bibinfo {year} {2017})}\BibitemShut {NoStop}%
\bibitem [{\citenamefont {Garrahan}\ and\ \citenamefont
  {Lesanovsky}(2010)}]{Garrahan2010}%
  \BibitemOpen
  \bibfield  {author} {\bibinfo {author} {\bibfnamefont {J.~P.}\ \bibnamefont
  {Garrahan}}\ and\ \bibinfo {author} {\bibfnamefont {I.}~\bibnamefont
  {Lesanovsky}},\ }\href {\doibase 10.1103/PhysRevLett.104.160601} {\bibfield
  {journal} {\bibinfo  {journal} {Phys. Rev. Lett.}\ }\textbf {\bibinfo
  {volume} {104}},\ \bibinfo {pages} {160601} (\bibinfo {year}
  {2010})}\BibitemShut {NoStop}%
\bibitem [{\citenamefont {Bobenko}\ \emph {et~al.}(1993)\citenamefont
  {Bobenko}, \citenamefont {Bordemann}, \citenamefont {Gunn},\ and\
  \citenamefont {Pinkall}}]{Bobenko1993}%
  \BibitemOpen
  \bibfield  {author} {\bibinfo {author} {\bibfnamefont {A.}~\bibnamefont
  {Bobenko}}, \bibinfo {author} {\bibfnamefont {M.}~\bibnamefont {Bordemann}},
  \bibinfo {author} {\bibfnamefont {C.}~\bibnamefont {Gunn}}, \ and\ \bibinfo
  {author} {\bibfnamefont {U.}~\bibnamefont {Pinkall}},\ }\href@noop {}
  {\bibfield  {journal} {\bibinfo  {journal} {Comm. Math. Phys}\ }\textbf
  {\bibinfo {volume} {158}},\ \bibinfo {pages} {127} (\bibinfo {year}
  {1993})}\BibitemShut {NoStop}%
\bibitem [{\citenamefont {Fredrickson}\ and\ \citenamefont
  {Andersen}(1984)}]{Fredrickson1984}%
  \BibitemOpen
  \bibfield  {author} {\bibinfo {author} {\bibfnamefont {G.~H.}\ \bibnamefont
  {Fredrickson}}\ and\ \bibinfo {author} {\bibfnamefont {H.~C.}\ \bibnamefont
  {Andersen}},\ }\href@noop {} {\bibfield  {journal} {\bibinfo  {journal}
  {Phys. Rev. Lett.}\ }\textbf {\bibinfo {volume} {53}},\ \bibinfo {pages}
  {1244} (\bibinfo {year} {1984})}\BibitemShut {NoStop}%
\bibitem [{\citenamefont {Garrahan}(2018)}]{Garrahan2018}%
  \BibitemOpen
  \bibfield  {author} {\bibinfo {author} {\bibfnamefont {J.~P.}\ \bibnamefont
  {Garrahan}},\ }\href {\doibase https://doi.org/10.1016/j.physa.2017.12.149}
  {\bibfield  {journal} {\bibinfo  {journal} {Physica A}\ }\textbf {\bibinfo
  {volume} {504}},\ \bibinfo {pages} {130} (\bibinfo {year}
  {2018})}\BibitemShut {NoStop}%
\bibitem [{\citenamefont {Takesue}(1987)}]{Takesue1987}%
  \BibitemOpen
  \bibfield  {author} {\bibinfo {author} {\bibfnamefont {S.}~\bibnamefont
  {Takesue}},\ }\href {\doibase 10.1103/PhysRevLett.59.2499} {\bibfield
  {journal} {\bibinfo  {journal} {Phys. Rev. Lett.}\ }\textbf {\bibinfo
  {volume} {59}},\ \bibinfo {pages} {2499} (\bibinfo {year}
  {1987})}\BibitemShut {NoStop}%
\bibitem [{\citenamefont {Prosen}\ and\ \citenamefont
  {Mej{\'\i}a-Monasterio}(2016)}]{Prosen2016}%
  \BibitemOpen
  \bibfield  {author} {\bibinfo {author} {\bibfnamefont {T.}~\bibnamefont
  {Prosen}}\ and\ \bibinfo {author} {\bibfnamefont {C.}~\bibnamefont
  {Mej{\'\i}a-Monasterio}},\ }\href
  {http://stacks.iop.org/1751-8121/49/i=18/a=185003} {\bibfield  {journal}
  {\bibinfo  {journal} {J. Phys. A}\ }\textbf {\bibinfo {volume} {49}},\
  \bibinfo {pages} {185003} (\bibinfo {year} {2016})}\BibitemShut {NoStop}%
\bibitem [{\citenamefont {Inoue}\ and\ \citenamefont
  {Takesue}(2018)}]{Inoue2018}%
  \BibitemOpen
  \bibfield  {author} {\bibinfo {author} {\bibfnamefont {A.}~\bibnamefont
  {Inoue}}\ and\ \bibinfo {author} {\bibfnamefont {S.}~\bibnamefont
  {Takesue}},\ }\href {http://stacks.iop.org/1751-8121/51/i=42/a=425001}
  {\bibfield  {journal} {\bibinfo  {journal} {J. Phys. A}\ }\textbf {\bibinfo
  {volume} {51}},\ \bibinfo {pages} {425001} (\bibinfo {year}
  {2018})}\BibitemShut {NoStop}%
\bibitem [{\citenamefont {Prosen}\ and\ \citenamefont {Bu{\v
  c}a}(2017)}]{Prosen2017}%
  \BibitemOpen
  \bibfield  {author} {\bibinfo {author} {\bibfnamefont {T.}~\bibnamefont
  {Prosen}}\ and\ \bibinfo {author} {\bibfnamefont {B.}~\bibnamefont {Bu{\v
  c}a}},\ }\href {http://stacks.iop.org/1751-8121/50/i=39/a=395002} {\bibfield
  {journal} {\bibinfo  {journal} {J. Phys. A}\ }\textbf {\bibinfo {volume}
  {50}},\ \bibinfo {pages} {395002} (\bibinfo {year} {2017})}\BibitemShut
  {NoStop}%
\bibitem [{\citenamefont {Klobas}\ \emph {et~al.}(2018)\citenamefont {Klobas},
  \citenamefont {Medenjak}, \citenamefont {Prosen},\ and\ \citenamefont
  {Vanicat}}]{Klobas2018}%
  \BibitemOpen
  \bibfield  {author} {\bibinfo {author} {\bibfnamefont {K.}~\bibnamefont
  {Klobas}}, \bibinfo {author} {\bibfnamefont {M.}~\bibnamefont {Medenjak}},
  \bibinfo {author} {\bibfnamefont {T.}~\bibnamefont {Prosen}}, \ and\ \bibinfo
  {author} {\bibfnamefont {M.}~\bibnamefont {Vanicat}},\ }\href@noop {}
  {\bibfield  {journal} {\bibinfo  {journal} {arXiv:1807.05000}\ } (\bibinfo
  {year} {2018})}\BibitemShut {NoStop}%
\bibitem [{\citenamefont {Derrida}\ and\ \citenamefont
  {Lebowitz}(1998)}]{Derrida1998}%
  \BibitemOpen
  \bibfield  {author} {\bibinfo {author} {\bibfnamefont {B.}~\bibnamefont
  {Derrida}}\ and\ \bibinfo {author} {\bibfnamefont {J.~L.}\ \bibnamefont
  {Lebowitz}},\ }\href@noop {} {\bibfield  {journal} {\bibinfo  {journal}
  {Phys. Rev. Lett.}\ }\textbf {\bibinfo {volume} {80}},\ \bibinfo {pages}
  {209} (\bibinfo {year} {1998})}\BibitemShut {NoStop}%
\bibitem [{\citenamefont {Prolhac}(2010)}]{Prolhac2010}%
  \BibitemOpen
  \bibfield  {author} {\bibinfo {author} {\bibfnamefont {S.}~\bibnamefont
  {Prolhac}},\ }\href@noop {} {\bibfield  {journal} {\bibinfo  {journal} {J.
  Phys. A}\ }\textbf {\bibinfo {volume} {43}},\ \bibinfo {pages} {105002}
  (\bibinfo {year} {2010})}\BibitemShut {NoStop}%
\bibitem [{\citenamefont {de~Gier}\ and\ \citenamefont
  {Essler}(2011)}]{Gier2011}%
  \BibitemOpen
  \bibfield  {author} {\bibinfo {author} {\bibfnamefont {J.}~\bibnamefont
  {de~Gier}}\ and\ \bibinfo {author} {\bibfnamefont {F.~H.~L.}\ \bibnamefont
  {Essler}},\ }\href {\doibase 10.1103/PhysRevLett.107.010602} {\bibfield
  {journal} {\bibinfo  {journal} {Phys. Rev. Lett.}\ }\textbf {\bibinfo
  {volume} {107}},\ \bibinfo {pages} {010602} (\bibinfo {year}
  {2011})}\BibitemShut {NoStop}%
\bibitem [{\citenamefont {Gorissen}\ \emph {et~al.}(2012)\citenamefont
  {Gorissen}, \citenamefont {Lazarescu}, \citenamefont {Mallick},\ and\
  \citenamefont {Vanderzande}}]{Gorissen2012}%
  \BibitemOpen
  \bibfield  {author} {\bibinfo {author} {\bibfnamefont {M.}~\bibnamefont
  {Gorissen}}, \bibinfo {author} {\bibfnamefont {A.}~\bibnamefont {Lazarescu}},
  \bibinfo {author} {\bibfnamefont {K.}~\bibnamefont {Mallick}}, \ and\
  \bibinfo {author} {\bibfnamefont {C.}~\bibnamefont {Vanderzande}},\ }\href
  {\doibase 10.1103/PhysRevLett.109.170601} {\bibfield  {journal} {\bibinfo
  {journal} {Phys. Rev. Lett.}\ }\textbf {\bibinfo {volume} {109}},\ \bibinfo
  {pages} {170601} (\bibinfo {year} {2012})}\BibitemShut {NoStop}%
\bibitem [{\citenamefont {Cramp{\'e}}\ \emph {et~al.}(2016)\citenamefont
  {Cramp{\'e}}, \citenamefont {Ragoucy}, \citenamefont {Rittenberg},\ and\
  \citenamefont {Vanicat}}]{Crampe2016}%
  \BibitemOpen
  \bibfield  {author} {\bibinfo {author} {\bibfnamefont {N.}~\bibnamefont
  {Cramp{\'e}}}, \bibinfo {author} {\bibfnamefont {E.}~\bibnamefont {Ragoucy}},
  \bibinfo {author} {\bibfnamefont {V.}~\bibnamefont {Rittenberg}}, \ and\
  \bibinfo {author} {\bibfnamefont {M.}~\bibnamefont {Vanicat}},\ }\href@noop
  {} {\bibfield  {journal} {\bibinfo  {journal} {Phys. Rev. E}\ }\textbf
  {\bibinfo {volume} {94}},\ \bibinfo {pages} {032102} (\bibinfo {year}
  {2016})}\BibitemShut {NoStop}%
\bibitem [{SM()}]{SM}%
  \BibitemOpen
  \href@noop {} {}\bibinfo {howpublished} {Supplemental Material}\BibitemShut
  {NoStop}%
\bibitem [{\citenamefont {Derrida}\ \emph {et~al.}(1993)\citenamefont
  {Derrida}, \citenamefont {Evans}, \citenamefont {Hakim},\ and\ \citenamefont
  {Pasquier}}]{Derrida1993}%
  \BibitemOpen
  \bibfield  {author} {\bibinfo {author} {\bibfnamefont {B.}~\bibnamefont
  {Derrida}}, \bibinfo {author} {\bibfnamefont {M.~R.}\ \bibnamefont {Evans}},
  \bibinfo {author} {\bibfnamefont {V.}~\bibnamefont {Hakim}}, \ and\ \bibinfo
  {author} {\bibfnamefont {V.}~\bibnamefont {Pasquier}},\ }\href@noop {}
  {\bibfield  {journal} {\bibinfo  {journal} {J. Phys. A}\ }\textbf {\bibinfo
  {volume} {26}},\ \bibinfo {pages} {1493} (\bibinfo {year}
  {1993})}\BibitemShut {NoStop}%
\bibitem [{\citenamefont {Vanicat}(2017)}]{Vanicat2017}%
  \BibitemOpen
  \bibfield  {author} {\bibinfo {author} {\bibfnamefont {M.}~\bibnamefont
  {Vanicat}},\ }\href@noop {} {\bibfield  {journal} {\bibinfo  {journal}
  {Journal of Statistical Physics}\ }\textbf {\bibinfo {volume} {166}},\
  \bibinfo {pages} {1129} (\bibinfo {year} {2017})}\BibitemShut {NoStop}%
\bibitem [{\citenamefont {Vanicat}(2018)}]{Vanicat2018}%
  \BibitemOpen
  \bibfield  {author} {\bibinfo {author} {\bibfnamefont {M.}~\bibnamefont
  {Vanicat}},\ }\href@noop {} {\bibfield  {journal} {\bibinfo  {journal}
  {Nuclear Physics B}\ }\textbf {\bibinfo {volume} {929}},\ \bibinfo {pages}
  {298} (\bibinfo {year} {2018})}\BibitemShut {NoStop}%
\bibitem [{\citenamefont {Finn}\ and\ \citenamefont
  {Vanicat}(2017)}]{Finn2017}%
  \BibitemOpen
  \bibfield  {author} {\bibinfo {author} {\bibfnamefont {C.}~\bibnamefont
  {Finn}}\ and\ \bibinfo {author} {\bibfnamefont {M.}~\bibnamefont {Vanicat}},\
  }\href {http://stacks.iop.org/1742-5468/2017/i=2/a=023102} {\bibfield
  {journal} {\bibinfo  {journal} {Journal of Statistical Mechanics: Theory and
  Experiment}\ }\textbf {\bibinfo {volume} {2017}},\ \bibinfo {pages} {023102}
  (\bibinfo {year} {2017})}\BibitemShut {NoStop}%
\bibitem [{\citenamefont {Cramp{\'e}}\ \emph {et~al.}(2011)\citenamefont
  {Cramp{\'e}}, \citenamefont {Ragoucy},\ and\ \citenamefont
  {Simon}}]{Crampe2011}%
  \BibitemOpen
  \bibfield  {author} {\bibinfo {author} {\bibfnamefont {N.}~\bibnamefont
  {Cramp{\'e}}}, \bibinfo {author} {\bibfnamefont {E.}~\bibnamefont {Ragoucy}},
  \ and\ \bibinfo {author} {\bibfnamefont {D.}~\bibnamefont {Simon}},\
  }\href@noop {} {\bibfield  {journal} {\bibinfo  {journal} {J. Phys. A}\
  }\textbf {\bibinfo {volume} {44}},\ \bibinfo {pages} {405003} (\bibinfo
  {year} {2011})}\BibitemShut {NoStop}%
\bibitem [{\citenamefont {Bodineau}\ \emph {et~al.}(2012)\citenamefont
  {Bodineau}, \citenamefont {Lecomte},\ and\ \citenamefont
  {Toninelli}}]{Bodineau2012}%
  \BibitemOpen
  \bibfield  {author} {\bibinfo {author} {\bibfnamefont {T.}~\bibnamefont
  {Bodineau}}, \bibinfo {author} {\bibfnamefont {V.}~\bibnamefont {Lecomte}}, \
  and\ \bibinfo {author} {\bibfnamefont {C.}~\bibnamefont {Toninelli}},\
  }\href@noop {} {\bibfield  {journal} {\bibinfo  {journal} {J. Stat. Phys.}\
  }\textbf {\bibinfo {volume} {147}},\ \bibinfo {pages} {1} (\bibinfo {year}
  {2012})}\BibitemShut {NoStop}%
\bibitem [{\citenamefont {Bodineau}\ and\ \citenamefont
  {Toninelli}(2012)}]{Bodineau2012b}%
  \BibitemOpen
  \bibfield  {author} {\bibinfo {author} {\bibfnamefont {T.}~\bibnamefont
  {Bodineau}}\ and\ \bibinfo {author} {\bibfnamefont {C.}~\bibnamefont
  {Toninelli}},\ }\href@noop {} {\bibfield  {journal} {\bibinfo  {journal}
  {Commun. Math. Phys.}\ }\textbf {\bibinfo {volume} {311}},\ \bibinfo {pages}
  {357} (\bibinfo {year} {2012})}\BibitemShut {NoStop}%
\bibitem [{\citenamefont {Nemoto}\ \emph {et~al.}(2017)\citenamefont {Nemoto},
  \citenamefont {Jack},\ and\ \citenamefont {Lecomte}}]{Nemoto2017}%
  \BibitemOpen
  \bibfield  {author} {\bibinfo {author} {\bibfnamefont {T.}~\bibnamefont
  {Nemoto}}, \bibinfo {author} {\bibfnamefont {R.~L.}\ \bibnamefont {Jack}}, \
  and\ \bibinfo {author} {\bibfnamefont {V.}~\bibnamefont {Lecomte}},\ }\href
  {\doibase 10.1103/PhysRevLett.118.115702} {\bibfield  {journal} {\bibinfo
  {journal} {Phys. Rev. Lett.}\ }\textbf {\bibinfo {volume} {118}},\ \bibinfo
  {pages} {115702} (\bibinfo {year} {2017})}\BibitemShut {NoStop}%
\bibitem [{\citenamefont {Jordan}\ and\ \citenamefont
  {Sukhorukov}(2004)}]{Jordan2004}%
  \BibitemOpen
  \bibfield  {author} {\bibinfo {author} {\bibfnamefont {A.~N.}\ \bibnamefont
  {Jordan}}\ and\ \bibinfo {author} {\bibfnamefont {E.~V.}\ \bibnamefont
  {Sukhorukov}},\ }\href {\doibase 10.1103/PhysRevLett.93.260604} {\bibfield
  {journal} {\bibinfo  {journal} {Phys. Rev. Lett.}\ }\textbf {\bibinfo
  {volume} {93}},\ \bibinfo {pages} {260604} (\bibinfo {year}
  {2004})}\BibitemShut {NoStop}%
\bibitem [{\citenamefont {Lambert}\ \emph {et~al.}(2015)\citenamefont
  {Lambert}, \citenamefont {Nori},\ and\ \citenamefont {Flindt}}]{Lambert2015}%
  \BibitemOpen
  \bibfield  {author} {\bibinfo {author} {\bibfnamefont {N.}~\bibnamefont
  {Lambert}}, \bibinfo {author} {\bibfnamefont {F.}~\bibnamefont {Nori}}, \
  and\ \bibinfo {author} {\bibfnamefont {C.}~\bibnamefont {Flindt}},\ }\href
  {\doibase 10.1103/PhysRevLett.115.216803} {\bibfield  {journal} {\bibinfo
  {journal} {Phys. Rev. Lett.}\ }\textbf {\bibinfo {volume} {115}},\ \bibinfo
  {pages} {216803} (\bibinfo {year} {2015})}\BibitemShut {NoStop}%
\bibitem [{\citenamefont {Brandner}\ \emph {et~al.}(2017)\citenamefont
  {Brandner}, \citenamefont {Maisi}, \citenamefont {Pekola}, \citenamefont
  {Garrahan},\ and\ \citenamefont {Flindt}}]{Brandner2017}%
  \BibitemOpen
  \bibfield  {author} {\bibinfo {author} {\bibfnamefont {K.}~\bibnamefont
  {Brandner}}, \bibinfo {author} {\bibfnamefont {V.~F.}\ \bibnamefont {Maisi}},
  \bibinfo {author} {\bibfnamefont {J.~P.}\ \bibnamefont {Pekola}}, \bibinfo
  {author} {\bibfnamefont {J.~P.}\ \bibnamefont {Garrahan}}, \ and\ \bibinfo
  {author} {\bibfnamefont {C.}~\bibnamefont {Flindt}},\ }\href {\doibase
  10.1103/PhysRevLett.118.180601} {\bibfield  {journal} {\bibinfo  {journal}
  {Phys. Rev. Lett.}\ }\textbf {\bibinfo {volume} {118}},\ \bibinfo {pages}
  {180601} (\bibinfo {year} {2017})}\BibitemShut {NoStop}%
\bibitem [{\citenamefont {Jack}\ and\ \citenamefont
  {Sollich}(2010)}]{Jack2010}%
  \BibitemOpen
  \bibfield  {author} {\bibinfo {author} {\bibfnamefont {R.~L.}\ \bibnamefont
  {Jack}}\ and\ \bibinfo {author} {\bibfnamefont {P.}~\bibnamefont {Sollich}},\
  }\href@noop {} {\bibfield  {journal} {\bibinfo  {journal} {Prog. Theor. Phys.
  Supp.}\ }\textbf {\bibinfo {volume} {184}},\ \bibinfo {pages} {304} (\bibinfo
  {year} {2010})}\BibitemShut {NoStop}%
\bibitem [{\citenamefont {Chetrite}\ and\ \citenamefont
  {Touchette}(2015{\natexlab{a}})}]{Chetrite2015}%
  \BibitemOpen
  \bibfield  {author} {\bibinfo {author} {\bibfnamefont {R.}~\bibnamefont
  {Chetrite}}\ and\ \bibinfo {author} {\bibfnamefont {H.}~\bibnamefont
  {Touchette}},\ }\href@noop {} {\bibfield  {journal} {\bibinfo  {journal}
  {Ann. Henri Poincar\'e}\ }\textbf {\bibinfo {volume} {16}},\ \bibinfo {pages}
  {2005} (\bibinfo {year} {2015}{\natexlab{a}})}\BibitemShut {NoStop}%
\bibitem [{\citenamefont {Borkar}\ \emph {et~al.}(2003)\citenamefont {Borkar},
  \citenamefont {Juneja},\ and\ \citenamefont {Kherani}}]{Borkar2003}%
  \BibitemOpen
  \bibfield  {author} {\bibinfo {author} {\bibfnamefont {V.~S.}\ \bibnamefont
  {Borkar}}, \bibinfo {author} {\bibfnamefont {S.}~\bibnamefont {Juneja}}, \
  and\ \bibinfo {author} {\bibfnamefont {A.~A.}\ \bibnamefont {Kherani}},\
  }\href {https://projecteuclid.org:443/euclid.cis/1119639799} {\bibfield
  {journal} {\bibinfo  {journal} {Commun. Inf. Syst.}\ }\textbf {\bibinfo
  {volume} {3}},\ \bibinfo {pages} {259} (\bibinfo {year} {2003})}\BibitemShut
  {NoStop}%
\bibitem [{\citenamefont {Chetrite}\ and\ \citenamefont
  {Touchette}(2015{\natexlab{b}})}]{Chetrite2015b}%
  \BibitemOpen
  \bibfield  {author} {\bibinfo {author} {\bibfnamefont {R.}~\bibnamefont
  {Chetrite}}\ and\ \bibinfo {author} {\bibfnamefont {H.}~\bibnamefont
  {Touchette}},\ }\href@noop {} {\bibfield  {journal} {\bibinfo  {journal} {J.
  Stat. Mech.}\ }\textbf {\bibinfo {volume} {2015}},\ \bibinfo {pages} {P12001}
  (\bibinfo {year} {2015}{\natexlab{b}})}\BibitemShut {NoStop}%
\bibitem [{\citenamefont {Jack}\ and\ \citenamefont
  {Sollich}(2015)}]{Jack2015b}%
  \BibitemOpen
  \bibfield  {author} {\bibinfo {author} {\bibfnamefont {R.~L.}\ \bibnamefont
  {Jack}}\ and\ \bibinfo {author} {\bibfnamefont {P.}~\bibnamefont {Sollich}},\
  }\href@noop {} {\bibfield  {journal} {\bibinfo  {journal} {Euro. Phys. J.
  Spec. Topics}\ }\textbf {\bibinfo {volume} {224}},\ \bibinfo {pages} {2351}
  (\bibinfo {year} {2015})}\BibitemShut {NoStop}%
\bibitem [{\citenamefont {Garrahan}(2016)}]{Garrahan2016}%
  \BibitemOpen
  \bibfield  {author} {\bibinfo {author} {\bibfnamefont {J.~P.}\ \bibnamefont
  {Garrahan}},\ }\href {http://stacks.iop.org/1742-5468/2016/i=7/a=073208}
  {\bibfield  {journal} {\bibinfo  {journal} {Journal of Statistical Mechanics:
  Theory and Experiment}\ }\textbf {\bibinfo {volume} {2016}},\ \bibinfo
  {pages} {073208} (\bibinfo {year} {2016})}\BibitemShut {NoStop}%
\bibitem [{\citenamefont {Carollo}\ \emph {et~al.}(2018)\citenamefont
  {Carollo}, \citenamefont {Garrahan}, \citenamefont {Lesanovsky},\ and\
  \citenamefont {P\'erez-Espigares}}]{Carollo2018}%
  \BibitemOpen
  \bibfield  {author} {\bibinfo {author} {\bibfnamefont {F.}~\bibnamefont
  {Carollo}}, \bibinfo {author} {\bibfnamefont {J.~P.}\ \bibnamefont
  {Garrahan}}, \bibinfo {author} {\bibfnamefont {I.}~\bibnamefont
  {Lesanovsky}}, \ and\ \bibinfo {author} {\bibfnamefont {C.}~\bibnamefont
  {P\'erez-Espigares}},\ }\href@noop {} {\bibfield  {journal} {\bibinfo
  {journal} {Phys. Rev. A}\ }\textbf {\bibinfo {volume} {98}},\ \bibinfo
  {pages} {010103} (\bibinfo {year} {2018})}\BibitemShut {NoStop}%
\bibitem [{\citenamefont {Derrida}\ and\ \citenamefont
  {Sadhu}(2018)}]{Derrida2018}%
  \BibitemOpen
  \bibfield  {author} {\bibinfo {author} {\bibfnamefont {B.}~\bibnamefont
  {Derrida}}\ and\ \bibinfo {author} {\bibfnamefont {T.}~\bibnamefont
  {Sadhu}},\ }\href@noop {} {\bibfield  {journal} {\bibinfo  {journal} {arXiv
  preprint arXiv:1807.06543}\ } (\bibinfo {year} {2018})}\BibitemShut {NoStop}%
\bibitem [{\citenamefont {Maes}\ and\ \citenamefont
  {Netocny}(2008)}]{Maes2008}%
  \BibitemOpen
  \bibfield  {author} {\bibinfo {author} {\bibfnamefont {C.}~\bibnamefont
  {Maes}}\ and\ \bibinfo {author} {\bibfnamefont {K.}~\bibnamefont {Netocny}},\
  }\href@noop {} {\bibfield  {journal} {\bibinfo  {journal} {Europhys. Lett.}\
  }\textbf {\bibinfo {volume} {82}},\ \bibinfo {pages} {30003} (\bibinfo {year}
  {2008})}\BibitemShut {NoStop}%
\bibitem [{\citenamefont {Bertini}\ \emph {et~al.}(2015)\citenamefont
  {Bertini}, \citenamefont {Faggionato},\ and\ \citenamefont
  {Gabrielli}}]{Bertini2015b}%
  \BibitemOpen
  \bibfield  {author} {\bibinfo {author} {\bibfnamefont {L.}~\bibnamefont
  {Bertini}}, \bibinfo {author} {\bibfnamefont {A.}~\bibnamefont {Faggionato}},
  \ and\ \bibinfo {author} {\bibfnamefont {D.}~\bibnamefont {Gabrielli}},\
  }\href@noop {} {\bibfield  {journal} {\bibinfo  {journal} {Stochastic
  Process. Appl.}\ }\textbf {\bibinfo {volume} {125}},\ \bibinfo {pages} {2786
  } (\bibinfo {year} {2015})}\BibitemShut {NoStop}%
\bibitem [{\citenamefont {Barato}\ and\ \citenamefont
  {Chetrite}(2015)}]{Barato2015b}%
  \BibitemOpen
  \bibfield  {author} {\bibinfo {author} {\bibfnamefont {A.~C.}\ \bibnamefont
  {Barato}}\ and\ \bibinfo {author} {\bibfnamefont {R.}~\bibnamefont
  {Chetrite}},\ }\href@noop {} {\bibfield  {journal} {\bibinfo  {journal} {J.
  Stat. Phys.}\ }\textbf {\bibinfo {volume} {160}},\ \bibinfo {pages} {1154}
  (\bibinfo {year} {2015})}\BibitemShut {NoStop}%
\bibitem [{\citenamefont {Hoppenau}\ \emph {et~al.}(2016)\citenamefont
  {Hoppenau}, \citenamefont {Nickelsen},\ and\ \citenamefont
  {Engel}}]{Hoppenau2016}%
  \BibitemOpen
  \bibfield  {author} {\bibinfo {author} {\bibfnamefont {J.}~\bibnamefont
  {Hoppenau}}, \bibinfo {author} {\bibfnamefont {D.}~\bibnamefont {Nickelsen}},
  \ and\ \bibinfo {author} {\bibfnamefont {A.}~\bibnamefont {Engel}},\
  }\href@noop {} {\bibfield  {journal} {\bibinfo  {journal} {New J. Phys.}\
  }\textbf {\bibinfo {volume} {18}},\ \bibinfo {pages} {083010} (\bibinfo
  {year} {2016})}\BibitemShut {NoStop}%
\bibitem [{\citenamefont {Bertini}\ \emph {et~al.}(2018)\citenamefont
  {Bertini}, \citenamefont {Chetrite}, \citenamefont {Faggionato},\ and\
  \citenamefont {Gabrielli}}]{Bertini2018}%
  \BibitemOpen
  \bibfield  {author} {\bibinfo {author} {\bibfnamefont {L.}~\bibnamefont
  {Bertini}}, \bibinfo {author} {\bibfnamefont {R.}~\bibnamefont {Chetrite}},
  \bibinfo {author} {\bibfnamefont {A.}~\bibnamefont {Faggionato}}, \ and\
  \bibinfo {author} {\bibfnamefont {D.}~\bibnamefont {Gabrielli}},\ }\href@noop
  {} {\bibfield  {journal} {\bibinfo  {journal} {Ann. Henri Poincar\'e}\
  }\textbf {\bibinfo {volume} {19}},\ \bibinfo {pages} {3197} (\bibinfo {year}
  {2018})}\BibitemShut {NoStop}%
\bibitem [{\citenamefont {Carollo}\ \emph {et~al.}(2017)\citenamefont
  {Carollo}, \citenamefont {Garrahan}, \citenamefont {Lesanovsky},\ and\
  \citenamefont {P{\'e}rez-Espigares}}]{Carollo2017}%
  \BibitemOpen
  \bibfield  {author} {\bibinfo {author} {\bibfnamefont {F.}~\bibnamefont
  {Carollo}}, \bibinfo {author} {\bibfnamefont {J.~P.}\ \bibnamefont
  {Garrahan}}, \bibinfo {author} {\bibfnamefont {I.}~\bibnamefont
  {Lesanovsky}}, \ and\ \bibinfo {author} {\bibfnamefont {C.}~\bibnamefont
  {P{\'e}rez-Espigares}},\ }\href@noop {} {\bibfield  {journal} {\bibinfo
  {journal} {Physical Review E}\ }\textbf {\bibinfo {volume} {96}},\ \bibinfo
  {pages} {052118} (\bibinfo {year} {2017})}\BibitemShut {NoStop}%
\end{thebibliography}%

\def \LL {{\cal L}}
\newcommand{\unit}[0]{\ensuremath{\text{\usefont{U}{bbold}{m}{n}1}}}

\onecolumngrid
\newpage

\renewcommand\thesection{S\arabic{section}}
\renewcommand\theequation{S\arabic{equation}}
\renewcommand\thefigure{S\arabic{figure}}
\setcounter{equation}{0}

\begin{center}
{\Large \emph{Supplemental material}: Exact large deviation statistics and trajectory phase transition of a deterministic boundary driven cellular automaton}
\end{center}

\section{Markovian dynamics}
 
\subsection{Markov operator}

We provide here details about the Markov matrix associated to the model \cite{Prosen2016,Prosen2017}. The matrix acts on the vector space $(\mathbb{R}^2)^{\otimes N}$
constructed as the tensor product of $N$ copies of the elementary space $\mathbb{R}^2$ (each associated to one site of the lattice). 
It can be written as the product of two operators, corresponding to the two half-time steps of the stochastic dynamics 
 \begin{equation}
  M = M_{\rm o}M_{\rm e}
 \end{equation}
with
\begin{equation}
 \begin{aligned}
  & M_{\rm e} = P_{123}P_{345}\cdots P_{N-3,N-2,N-1} R_{N-1,N} \\
  & M_{\rm o} = L_{12} P_{234}P_{456} \cdots P_{N-2,N-1,N}
 \end{aligned}
\end{equation}
We have used the notation 
\begin{equation}
 P_{i-1,i,i+1} = \unit_2^{\otimes (i-2)} \otimes P \otimes \unit_2^{\otimes (N-i-1)}, \quad
 L_{12} = L \otimes \unit_2^{\otimes (N-2)}, \quad R_{N-1,N} = \unit_2^{\otimes (N-2)} \otimes R
\end{equation}
where $\unit_2$ is the identity matrix acting on the elementary space $\mathbb{R}^2$.
$P$ is a $8\times8$ permutation matrix encoding the deterministic dynamics in the bulk and 
acting on the space $(\mathbb{R}^2)^{\otimes 3}$ (\textit{i.e.} it acts on three sites of the lattice)
$L$ and $R$ are $4\times4$ matrices encoding the stochastic dynamics at the boundaries and acting on the space 
$(\mathbb{R}^2)^{\otimes 2}$ (\textit{i.e.} they act on two sites of the lattice).
These local matrices read explicitly \cite{Prosen2016,Prosen2017}
\begin{equation}
 P = \begin{pmatrix}
      1 & 0 & 0 & 0 & 0 & 0 & 0 & 0 \\ 
      0 & 0 & 0 & 1 & 0 & 0 & 0 & 0 \\
      0 & 0 & 1 & 0 & 0 & 0 & 0 & 0 \\
      0 & 1 & 0 & 0 & 0 & 0 & 0 & 0 \\
      0 & 0 & 0 & 0 & 0 & 0 & 1 & 0 \\
      0 & 0 & 0 & 0 & 0 & 0 & 0 & 1 \\
      0 & 0 & 0 & 0 & 1 & 0 & 0 & 0 \\
      0 & 0 & 0 & 0 & 0 & 1 & 0 & 0
     \end{pmatrix}
\end{equation}
and
\begin{equation}
 L = \begin{pmatrix}
          \alpha & 0 & \alpha & 0 \\
          0 & \beta & 0 & \beta \\
          1-\alpha & 0 & 1-\alpha & 0 \\
          0 & 1-\beta & 0 & 1-\beta
         \end{pmatrix}, \qquad 
 R = \begin{pmatrix}
          \gamma & \gamma & 0 & 0 \\
          1-\gamma & 1-\gamma & 0 & 0 \\
          0 & 0 & \delta & \delta \\
          0 & 0 & 1-\delta & 1-\delta
         \end{pmatrix}.
\end{equation}

\subsection{Tilted operator}

In order to study the statistics of the time-integrated observables we introduce a deformation of the Markov operator
\begin{equation}
 M(s) = M_{\rm o} \, G(s) \, M_{\rm e} \, F(s).
\end{equation}
It will be convenient for the following, when deriving the cancellation scheme for the matrix product expression of the leading eigenvector, to
introduce the following diagonal matrices
\begin{equation}
 F^{(i)} = \begin{pmatrix}
            f_{0,0}^{(i)} & 0 & 0 & 0 \\
            0 & f_{0,1}^{(i)} & 0 & 0 \\
            0 & 0 & f_{1,0}^{(i)} & 0 \\
            0 & 0 & 0 & f_{1,1}^{(i)}
           \end{pmatrix}, \qquad 
 G^{(i)} = \begin{pmatrix}
            g_{0,0}^{(i)} & 0 & 0 & 0 \\
            0 & g_{0,1}^{(i)} & 0 & 0 \\
            0 & 0 & g_{1,0}^{(i)} & 0 \\
            0 & 0 & 0 & g_{1,1}^{(i)}
           \end{pmatrix}
\end{equation}
Then the extensive tilt operators $F(s)$ and $G(s)$ can be expressed as
\begin{equation}
 F(s) = F^{(1)}_{12}F^{(2)}_{23}F^{(3)}_{34} \dots F^{(N-1)}_{N-1,N} \quad \mbox{and} \quad G(s) = G^{(1)}_{12}G^{(2)}_{23}G^{(3)}_{34} \dots G^{(N-1)}_{N-1,N}
\end{equation}
where we have again used subscript index to denote the sites of the lattice on which the operators are acting non-trivially.
A direct computation shows that we can distribute those local operators $F^{(j)}$ and $G^{(j)}$ on the local matrices encoding the dynamics of the model
\begin{equation} \label{eq:product_tilted_Markov}
\begin{aligned}
 & M_{\rm e} \, F(s) = \tilde P^{(1)}_{123}\tilde P^{(3)}_{345}\cdots \tilde P^{(N-3)}_{N-3,N-2,N-1}\tilde R^{(N-1)}_{N-1,N} \\
 & M_{\rm o} \, G(s) = \tilde L^{(1)}_{12}\tilde P^{(2)}_{234}\tilde P^{(4)}_{456} \cdots \tilde P^{(N-2)}_{N-2,N-1,N}
 \end{aligned}
\end{equation}
where 
\begin{equation}
 \tilde L^{(1)} = L G^{(1)}, \qquad \tilde R^{(N-1)} = R F^{(N-1)},
\end{equation}
and, for $i$ even,
\begin{equation}
\tilde P^{(i-1)} =  P F^{(i-1)}_{12}F^{(i)}_{23}, \qquad \tilde P^{(i)} =  P G^{(i)}_{12}G^{(i+1)}_{23}.
\end{equation}
The structure of the tilted Markov operator given in \eqref{eq:product_tilted_Markov} as the product of local operators will be very convenient to
prove the matrix ansatz construction of the leading eigenvector.

\section{Matrix product expression of the leading eigenvector}

This section is devoted to the construction of the leading eigenvector of the tilted operator in a matrix product form.
Recall that our strategy is to look for a pair of vectors $\bm{p}$ and $\bm{p'}$ such that
\begin{equation} \label{eq:pair_eigenvectors}
 M_{\rm e}F(s) \bm{p} = \lambda_{R} \bm{p'} \quad \mbox{and} \quad M_{\rm o}G(s) \bm{p'} = \lambda_{L} \bm{p}
 \end{equation}
and such that $\lambda(s) = \lambda_{R}\lambda_{L}$ is the dominant eigenvalue of the tilted Markov operator $M(s)$. 
This would indeed imply $M(s) \bm{p} = \lambda(s) \bm{p}$.

We are going to explain the exact construction of these vectors in a matrix product form
\begin{equation} \label{eq:matrix_ansatz}
  \begin{aligned}
  & p_{n_1,n_2,\dots,n_N} =  \bra{l_{n_1}} W_{n_2}^{(2)}W_{n_3}^{(3)} \cdots W_{n_{N-3}}^{(N-3)}W_{n_{N-2}}^{(N-2)}\ket{r_{n_{N-1}n_N}} \\
  & p'_{n_1,n_2,\dots,n_N} =  \bra{l'_{n_1 n_2}} V_{n_3}^{(3)}V_{n_4}^{(4)} \cdots V_{n_{N-2}}^{(N-2)}V_{n_{N-1}}^{(N-1)}\ket{r'_{n_N}}.
  \end{aligned}
 \end{equation}
 where $W_{n}^{(k)},V_{n}^{(k)}$, $n\in\{0,1\}$, are $3 \times 3$ matrices acting in an auxiliary space, 
and where $\bra{l_{n}},\bra{l'_{n,n'}}$, $n,n'\in\{0,1\}$, are row vectors in the auxiliary space (with $3$ entries) and 
 $\ket{r'_{n}},\ket{r_{n,n'}}$, $n,n'\in\{0,1\}$, are column vectors in the auxiliary space (with $3$ entries).
 The proof of the matrix product structure of this pair of vectors will be done in two steps. First we will present a 
 cancellation scheme, which shows that the pair of relations \eqref{eq:pair_eigenvectors} can be obtained if the auxiliary matrices and vectors arising
 in the ansatz \eqref{eq:matrix_ansatz} satisfy very simple algebraic relations. Second we will construct explicit solutions of these 
 algebraic relations using $3\times3$ matrices and $3$-components vectors.

\subsection{Cancellation scheme}

In order to prove efficiently our exact construction and to present the cancellation scheme, we introduce the vectors with matrix components 
\begin{equation}
  \bm{W}^{(i)} = \begin{pmatrix}
                  W_0^{(i)} \\ W_1^{(i)}
                 \end{pmatrix}, \qquad 
  \bm{V}^{(i)} = \begin{pmatrix}
                  V_0^{(i)} \\ V_1^{(i)}
                 \end{pmatrix}, \qquad 
  \bm{X}^{(i)} = \begin{pmatrix}
                  X_0^{(i)} \\ X_1^{(i)}
                 \end{pmatrix}
 \end{equation}
 with $X_{n}^{(j)}$, $n\in\{0,1\}$, $i\in\{2,\ldots,N-1\}$ being $3 \times 3$ matrices acting in the auxiliary space, which will play an important role in the proof of the 
 matrix ansatz.
 We also introduce
 \begin{equation}
  \bra{\bm{l}}= \begin{pmatrix}
                  \bra{l_0} \\ \bra{l_1}
                 \end{pmatrix}, \qquad 
  \bra{\bm{l'}}= \begin{pmatrix}
                  \bra{l'_{00}} \\ \bra{l'_{01}} \\ \bra{l'_{10}} \\ \bra{l'_{11}}
                 \end{pmatrix}, \qquad
  \ket{\bm{r'}}= \begin{pmatrix}
                  \ket{r'_0} \\ \ket{r'_1}
                 \end{pmatrix}, \qquad 
  \ket{\bm{r}}= \begin{pmatrix}
                  \ket{r_{00}} \\ \ket{r_{01}} \\ \ket{r_{10}} \\ \ket{r_{11}}
                 \end{pmatrix},
 \end{equation}
Using those new objects we can rewrite concisely the pair of vectors 
\begin{equation} \label{eq:tensor_prod_ansatz}
  \begin{aligned}
  & \bm{p} =  \bra{\bm{l}_1} \bm{W}_2^{(2)}\bm{W}_3^{(3)} \cdots \bm{W}_{N-3}^{(N-3)}\bm{W}_{N-2}^{(N-2)}\ket{\bm{r}_{N-1,N}} , \\
  & \bm{p'} =  \bra{\bm{l}_{12}'} \bm{V}_3^{(3)}\bm{V}_4^{(4)} \cdots \bm{V}_{N-2}^{(N-2)}\bm{V}_{N-1}^{(N-1)}\ket{\bm{r'}_{N}} .
  \end{aligned}
 \end{equation} 
 We used once again tensor product notations: the subscript indices stand for the tensor product component (\textit{i.e.} the site of the lattice)
 the vectors are belonging to. 
 As an illustration of our convention, we have for instance
 \begin{equation}
  \bm{W}_{i}^{(i)}\bm{W}_{i+1}^{(i+1)} = \begin{pmatrix}
                                          W^{(i)}_0 W^{(i+1)}_0 \\
                                          W^{(i)}_0 W^{(i+1)}_1 \\
                                          W^{(i)}_1 W^{(i+1)}_0 \\
                                          W^{(i)}_1 W^{(i+1)}_1 
                                         \end{pmatrix}, \qquad
  \bm{V}_{N-1}^{(N-1)}\ket{\bm{r'}_{N}} = \begin{pmatrix}
                                          V^{(N-1)}_0 \ket{r'_0} \\
                                          V^{(N-1)}_0 \ket{r'_1} \\
                                          V^{(N-1)}_1 \ket{r'_0} \\
                                          V^{(N-1)}_1 \ket{r'_1} 
                                         \end{pmatrix} .
 \end{equation}
 The tensor product notation \eqref{eq:tensor_prod_ansatz} will be very convenient to present the cancellation mechanism.  
 
 The cancellation scheme consists of two key relations. First, the inhomogeneous bulk relations, written for even site index $j=2i$:
\begin{equation} \label{eq:rel_bulk_comp}
\begin{aligned}
 f^{(j-1)}_{nn'} f^{(j)}_{n'n'\!'} W^{(j-1)}_n W^{(j)}_{n'} X^{(j+1)}_{n'\!'} &= 
X^{(j-1)}_{n} V^{(j)}_{\chi(nn'n'\!')} V^{(j+1)}_{n'\!'}, \\
 g^{(j-2)}_{nn'} g^{(j-1)}_{n'n'\!'} X^{(j-2)}_n V^{(j-1)}_{n'} V^{(j)}_{n'\!'} &= W^{(j-2)}_n W^{(j-1)}_{\chi(nn'n'\!')} X^{(j)}_{n'\!'}.
\end{aligned}
\end{equation}
Second, the boundary equations:
\begin{equation} \label{eq:rel_boundaries_comp}
\begin{aligned}
 f^{(1)}_{nn'}f^{(2)}_{n'n'\!'} \bra{l_n}W^{(2)}_{n'} X^{(3)}_{n'\!'} &= \bra{l'_{n\chi(nn'n'\!')}}V^{(3)}_{n'\!'},  \\
 \sum_{m,m'=0,1} R_{nn'}^{mm'}f^{(N-1)}_{mm'}\ket{r_{mm'}} &= \lambda_{\rm R} X^{(N-1)}_{n}\ket{r'_{n'}}, \\
 \sum_{m,m'=0,1} L_{nn'}^{mm'} g^{(1)}_{mm'} \bra{l'_{mm'}} &= \lambda_{\rm L} \bra{l_n}X^{(2)}_{n'},\\
 g^{(N-2)}_{nn'}\!g^{(N-1)}_{n'n'\!'} X^{(N-2)}_{n} V^{(N-1)}_{n'}\!\ket{r'_{n'\!'}} &= W^{(N-2)}_{n}\!\ket{r_{\!\chi(nn'n'\!')n'\!'}}.
\end{aligned}
\end{equation}
  Using tensor product notations, the inhomogeneous bulk equation can be concisely rewritten, for even $j$:
  \begin{equation} \label{eq:rel_bulk}
  \begin{aligned}
   \tilde P_{j-1,j,j+1}^{(j-1)} \bm{W}_{j-1}^{(j-1)} \bm{W}_j^{(j)} \bm{X}_{j+1}^{(j+1)} &=
   \bm{X}_{j-1}^{(j-1)} \bm{V}_j^{(j)} \bm{V}_{j+1}^{(j+1)} \\
   \tilde P_{j-2,j-1,j}^{(j-2)} \bm{X}_{j-2}^{(j-2)} \bm{V}_{j-1}^{(j-1)} \bm{V}_j^{(j)} &= 
   \bm{W}_{j-2}^{(j-2)} \bm{W}_{j-1}^{(j-1)} \bm{X}_j^{(j)}.
   \end{aligned}
  \end{equation}
 Similarly, the boundary equations can be rewritten, as
  \begin{equation} \label{eq:rel_left}
   \tilde L^{(1)}_{12} \bra{\bm{l'}_{12}} = \lambda_{L} \bra{\bm{l}_1} \bm{X}_2^{(2)}, \qquad 
   \tilde P^{(1)}_{123} \bra{\bm{l}_1}\bm{W}_2^{(2)}\bm{X}_3^{(3)} = \bra{\bm{l'}_{12}} \bm{V}_3^{(3)},
  \end{equation}
  for the left boundary, and
 \begin{equation} \label{eq:rel_right}
  \tilde R^{(N-1)}_{N-1,N} \ket{\bm{r}_{N-1,N}} = \lambda_{R} \bm{X}_{N-1}^{(N-1)}\ket{\bm{r'}_N}, \qquad 
  \tilde P^{(N-2)}_{N-2,N-1,N} \bm{X}_{N-2}^{(N-2)}\bm{V}_{N-1}^{(N-1)}\ket{\bm{r'}_N} = \bm{W}_{N-2}^{(N-2)}\ket{\bm{r}_{N-1,N}},
 \end{equation}
 for the right boundary.
 A direct computation shows that the algebraic relations \eqref{eq:rel_bulk}, \eqref{eq:rel_left} and \eqref{eq:rel_right} imply the 
 eigenvalue equations \eqref{eq:pair_eigenvectors}. We illustrate the mechanism in the case of a lattice of length $N=6$
 \begin{equation}
  \begin{aligned}
   M_{\rm e}F(s) \bm{p} &= \tilde P^{(1)}_{123}\tilde P^{(3)}_{345}\tilde R^{(5)}_{56} 
                         \bra{\bm{l}_1} \bm{W}_2^{(2)}\bm{W}_3^{(3)}\bm{W}_{4}^{(4)}\ket{\bm{r}_{5,6}} \\
                        &= \lambda_{R} \tilde P^{(1)}_{123}\tilde P^{(3)}_{345} 
                        \bra{\bm{l}_1} \bm{W}_2^{(2)}\bm{W}_3^{(3)}\bm{W}_{4}^{(4)}\bm{X}_{5}^{(5)}\ket{\bm{r'}_6} \\
                        &= \lambda_{R} \tilde P^{(1)}_{123}
                        \bra{\bm{l}_1} \bm{W}_2^{(2)}\bm{X}_3^{(3)}\bm{V}_{4}^{(4)}\bm{V}_{5}^{(5)}\ket{\bm{r'}_6} \\
                        &= \lambda_{R} \bra{\bm{l'}_{12}} \bm{V}_3^{(3)}\bm{V}_{4}^{(4)}\bm{V}_{5}^{(5)}\ket{\bm{r'}_6} \\
                        &= \lambda_{R} \bm{p'}.
  \end{aligned}
 \end{equation}
The relation $M_{\rm o}G(s) \bm{p'} = \lambda_{L} \bm{p}$ is established in a very similar way. Note that the value of the parameters
$\lambda_{L}$ and $\lambda_{R}$ are still free at this stage. They will be fixed when solving explicitly the algebraic relations  
\eqref{eq:rel_bulk}, \eqref{eq:rel_left} and \eqref{eq:rel_right} with finite dimensional matrices and vectors, and will provide
the eigenvalue $\lambda(s)=\lambda_{L}\lambda_{R}$.
 
\subsection{Solution to the inhomogeneous bulk algebra}

 We proceed to show how to explicitly solve the bulk relations \eqref{eq:rel_bulk}. We start by the matrix ansatz
 \begin{equation}
  W_0^{(i)} = \begin{pmatrix}
               1 & 0 & 0 \\
               w_1^{(i)} & 0 & 0 \\
               1 & 0 & 0
              \end{pmatrix}, \qquad 
  W_1^{(i)} = \begin{pmatrix}
               0 & w_2^{(i)} & 0 \\
               0 & 0 & 1 \\
               0 & 0 & w_3^{(i)}
              \end{pmatrix},
 \end{equation}
 
 \begin{equation}
  V_0^{(i)} = \begin{pmatrix}
               1 & 0 & 0 \\
               v_1^{(i)} & 0 & 0 \\
               1 & 0 & 0
              \end{pmatrix}, \qquad 
  V_1^{(i)} = \begin{pmatrix}
               0 & v_2^{(i)} & 0 \\
               0 & 0 & 1 \\
               0 & 0 & v_3^{(i)}
              \end{pmatrix},
 \end{equation}

 \begin{equation}
  X_0^{(i)} = \begin{pmatrix}
               x_1^{(i)} & 0 & 0 \\
               x_2^{(i)} & 0 & 0 \\
               x_3^{(i)} & 0 & 0
              \end{pmatrix}, \qquad 
  X_1^{(i)} = \begin{pmatrix}
               0 & 0 & x_4^{(i)} \\
               0 & x_5^{(i)} & 0 \\
               0 & x_6^{(i)} & 0
              \end{pmatrix},
 \end{equation}
 We stress here that these matrices are acting on a 3-dimensional auxiliary space which is different from the physical space of local configuration
 $\mathbb{R}^2$. This 3-dimensional vector space is traced out in the matrix product expression \eqref{eq:matrix_ansatz}
 thanks to the action of the boundary vectors.
 
 Plugging the matrix ansatz into \eqref{eq:rel_bulk}, we end up with the following two-parameter $(\rho,\kappa)$ solution
 \begin{equation}
  \begin{aligned}
  & w_1^{(2k)} = \kappa b^{(k)} \frac{t_{10}^{(2k)}}{u_{10}^{(2k)}}, &\quad &  w_2^{(2k)} = \kappa b^{(k)}, &\quad &  w_3^{(2k)} = \rho c^{(k)}, \\
  & w_1^{(2k+1)} = \rho c^{(k)} \frac{1}{u_{00}^{(2k)}}, &\quad & w_2^{(2k+1)} = \rho c^{(k+1)}, &\quad &  w_3^{(2k+1)} = \kappa b^{(k)}t_{00}^{(2k)},
  \end{aligned}
  \end{equation}
  \begin{equation}
  \begin{aligned}
  & v_1^{(2k)} = \rho c^{(k)}t_{10}^{(2k)}, &\quad & v_2^{(2k)} = \rho c^{(k)}\frac{1}{z^{(2k)}}, &\quad &  v_3^{(2k)} = \kappa b^{(k)} \frac{1}{u_{10}^{(2k)}}, \\
  & v_1^{(2k+1)} = \kappa b^{(k)}t_{00}^{(2k)}u_{10}^{(2k+1)}, &\quad & v_2^{(2k+1)} = \kappa b^{(k)}t_{00}^{(2k)}t_{10}^{(2k+1)}, &\quad & v_3^{(2k+1)} = \rho c^{(k)} \frac{1}{u_{00}^{(2k)}u_{10}^{(2k+1)}},
  \end{aligned}
 \end{equation}
and 
\begin{equation}
  \begin{aligned}
  & x_1^{(2k)} = \prod_{l=1}^{2k-1} g_{00}^{(l)} , &\quad & x_2^{(2k)} = x_1^{(2k)}w_1^{(2k)}, &\quad & x_3^{(2k)} = x_1^{(2k)}, \\ 
  & x_1^{(2k+1)} = \prod_{l=1}^{2k} \frac{1}{f_{00}^{(l)}}, &\quad & x_2^{(2k+1)} = x_1^{(2k+1)}w_1^{(2k+1)}, &\quad & x_3^{(2k+1)} = x_1^{(2k+1)},
  \end{aligned}
  \end{equation}
  \begin{equation}
  \begin{aligned}
  & x_4^{(2k)} = x_1^{(2k)}\kappa b^{(k)}\frac{g_{00}^{(2k)}}{g_{10}^{(2k)}}, &\quad &   
  x_5^{(2k)} = x_1^{(2k)}\frac{1}{z^{(2k)}}\frac{g_{01}^{(2k-1)}}{g_{00}^{(2k-1)}}, &\quad &   
  x_6^{(2k)} = x_1^{(2k)}\frac{\rho c^{(k)}}{z^{(2k)}}\frac{g_{01}^{(2k-1)}}{g_{00}^{(2k-1)}}, \\
  & x_4^{(2k+1)} = x_1^{(2k+1)}\frac{\rho c^{(k)}}{z^{(2k)}}\frac{f_{00}^{(2k)}}{f_{01}^{(2k)}}, &\quad &
  x_5^{(2k+1)} = x_1^{(2k+1)}\frac{f_{10}^{(2k+1)}}{f_{00}^{(2k+1)}}, &\quad &
  x_6^{(2k+1)} = x_1^{(2k+1)}\kappa b^{(k)}y^{(2k)}\frac{f_{10}^{(2k)}}{f_{11}^{(2k)}},
  \end{aligned}
 \end{equation}
 where 
 \begin{equation}
  \begin{aligned}
   & y^{(i)} = \frac{f_{01}^{(i-1)}f_{11}^{(i)}f_{10}^{(i+1)}}{f_{00}^{(i-1)}f_{00}^{(i)}f_{00}^{(i+1)}}, &\quad & 
   z^{(i)} = \frac{g_{01}^{(i-1)}g_{11}^{(i)}g_{10}^{(i+1)}}{g_{00}^{(i-1)}g_{00}^{(i)}g_{00}^{(i+1)}} \\
   & t_{nn'}^{(i)} = \frac{f_{n1}^{(i-1)}f_{1n'}^{(i)}}{f_{n0}^{(i-1)}f_{0n'}^{(i)}}, &\quad &
   u_{nn'}^{(i)} = \frac{g_{n1}^{(i-1)}g_{1n'}^{(i)}}{g_{n0}^{(i-1)}g_{0n'}^{(i)}}, \\
   & b^{(i)} = \prod_{k=0}^{i-1} y^{(2k)}z^{(2k+1)}, &\quad & c^{(i)} = \prod_{k=0}^{i-1} \frac{1}{y^{(2k+1)}z^{(2k)}},
  \end{aligned}
 \end{equation}
with the convention that $f_{nn'}^{(-1)}=f_{nn'}^{(0)}=1$ and $g_{nn'}^{(-1)}=g_{nn'}^{(0)}=1$.

Note that the 3-dimensional auxiliary space, arising in the explicit representation of the inhomogeneous bulk algebra, plays essentially the role
of detecting the left and right moving solitons in a given configuration. More precisely, in order to detect if a soliton is located at a given site 
of the lattice, it is necessary to know the content of the two neighboring sites (the number of left and right moving solitons is indeed a locally
conserved charge with support of size three).  When scanning a configuration from left to right,
the auxiliary space allows to store the information about the content of the two last sites and thus permits the detection of the solitons.
The $3\times 3$ matrices associate a weight (that could be thought as a momentum) to each soliton, which depends on its position on the lattice
and on its direction of propagation. We will investigate further in future work the connection between the conserved charges of the model and 
the matrix product construction of its eigenvectors.
 
\subsection{Solution to the boundary relations}

We first recall the convenient notations 
$\alpha'=\alpha+\widetilde{\alpha}$, $\beta'=\beta+\widetilde{\beta}$, $\gamma'=\gamma+\widetilde{\gamma}$, $\delta'=\delta+\widetilde{\delta}$,
 \begin{equation}
  \widetilde\alpha = \frac{f_{10}^{(1)}g_{11}^{(1)}}{f_{00}^{(1)}g_{01}^{(1)}}(1-\alpha), \quad 
  \widetilde\beta = \frac{f_{11}^{(1)}g_{10}^{(1)}}{f_{01}^{(1)}g_{00}^{(1)}}(1-\beta), \quad
  \widetilde\gamma = \frac{f_{11}^{(N-1)}g_{01}^{(N-1)}}{f_{10}^{(N-1)}g_{00}^{(N-1)}}(1-\gamma), \quad 
  \widetilde\delta = \frac{f_{01}^{(N-1)}g_{11}^{(N-1)}}{f_{00}^{(N-1)}g_{10}^{(N-1)}}(1-\delta),
 \end{equation}
It will be also useful to introduce the notations 
\begin{equation}
\begin{aligned}
 & \mathfrak{f}_N = \prod_{i=1}^{N-1} f_{00}^{(i)}, \qquad  \mathfrak{g}_N = \prod_{i=1}^{N-1} g_{00}^{(i)}, \qquad 
 \mathfrak{a}_N = \mathfrak{f}_N\mathfrak{g}_N, \\
 & \mathfrak{b}_N = b^{(N/2+1)}=\prod_{i=1}^{N/2} \frac{f_{01}^{(2i-1)}f_{10}^{(2i-1)}g_{11}^{(2i-1)}}{\big(f_{00}^{(2i-1)}\big)^2 g_{00}^{(2i-1)}}
  \prod_{i=1}^{N/2-1} \frac{g_{01}^{(2i)}g_{10}^{(2i)}f_{11}^{(2i)}}{\big(g_{00}^{(2i)}\big)^2 f_{00}^{(2i)}}, \\
 & \mathfrak{c}_N = \frac{1}{c^{(N/2+1)}}=\prod_{i=1}^{N/2} \frac{g_{01}^{(2i-1)}g_{10}^{(2i-1)}f_{11}^{(2i-1)}}{\big(g_{00}^{(2i-1)}\big)^2 f_{00}^{(2i-1)}}
  \prod_{k=1}^{N/2-1} \frac{f_{01}^{(2i)}f_{10}^{(2i)}g_{11}^{(2i)}}{\big(f_{00}^{(2i)}\big)^2 g_{00}^{(2i)}},
\end{aligned}
\end{equation}
with the convention that $f_{nn'}^{(N)}=f_{nn'}^{(N+1)}=1$ and $g_{nn'}^{(N)}=g_{nn'}^{(N+1)}=1$.

The resolution of the left boundary relation \eqref{eq:rel_left} fixes the value of the parameters $\rho$ and $\kappa$ appearing in the matrices 
$W^{(i)}_n$ and $V^{(i)}_n$
 \begin{equation} \label{eq:constraint_left}
  \rho = \frac{\lambda_{L}(\lambda_{L}-\alpha)}{(1-\beta)\alpha'}, \qquad 
  \kappa = \frac{(1-\alpha)(\widetilde\alpha \widetilde\beta -\alpha\beta+\beta\lambda_{L})\beta'}{\lambda_{L}^2\widetilde\alpha \widetilde\beta},
 \end{equation}
 and imposes the following explicit expression for the left boundary vectors
 \begin{equation}
  \begin{aligned}
  & \bra{l_{0}}= \Big( 1, \frac{\beta\widetilde\beta\lambda_L(\varepsilon-\widetilde\beta)+\rho(1-\beta)(\beta\widetilde\beta\lambda_L-\varepsilon\alpha\beta')}{z^{(0)}y^{(1)}\beta\widetilde\beta\lambda_L(1-\rho\kappa)(1-\beta)} , 
  -\frac{\kappa\beta\widetilde\beta\lambda_L(\varepsilon-\widetilde\beta)+(1-\beta)(\beta\widetilde\beta\lambda_L-\varepsilon\alpha\beta')}{\beta\widetilde\beta\lambda_L(1-\rho\kappa)(1-\beta)} \Big) \\
  & \bra{l_{1}}= \Big( \frac{1-\beta}{\beta} , \frac{\widetilde\beta\lambda_L(\varepsilon-\widetilde\beta)+\rho(1-\beta)\widetilde\beta\lambda_L-\rho(1-\alpha)\beta'\varepsilon}{z^{(0)}y^{(1)}\beta\widetilde\beta\lambda_L(1-\rho\kappa)} , 
  -\frac{\kappa\widetilde\beta\lambda_L(\varepsilon-\widetilde\beta)+(1-\beta)\widetilde\beta\lambda_L-(1-\alpha)\beta'\varepsilon}{\beta\widetilde\beta\lambda_L(1-\rho\kappa)} \Big) \\
  & \bra{l'_{00}}= \Big( \frac{\varepsilon}{\widetilde\beta} , \frac{\lambda_L(\alpha-\lambda_L)}{z^{(2)}t_{00}^{(2)}\beta\alpha'} , \frac{\widetilde\beta}{\beta}\Big), \qquad 
   \bra{l'_{01}}= \Big( 0 , \frac{\lambda_L(\lambda_L-\alpha)\beta'}{z^{(0)}y^{(1)}z^{(2)}\beta(1-\beta)\alpha'}, 
   \frac{(\alpha-\lambda_L)(\widetilde\alpha\widetilde\beta-\alpha\beta)\beta'}{\lambda_L\beta\widetilde\beta \alpha'} \Big) \\
  & \bra{l'_{10}}= \Big( \frac{\varepsilon}{z^{(0)}\beta} , \frac{\lambda_L(\lambda_L-\alpha)}{z^{(0)}z^{(2)}t_{00}^{(2)}\beta \alpha'} , -\frac{\widetilde\beta}{z^{(0)}\beta} \Big),  \qquad 
   \bra{l'_{11}}= \Big( 0 , 0 , \frac{\varepsilon \beta'(1-\alpha)t_{00}^{(1)}}{\lambda_L\beta\widetilde\beta} \Big)
  \end{aligned}
 \end{equation}
where we have introduced the parameter
\begin{equation}
\varepsilon = \frac{\widetilde\alpha\widetilde\beta-\alpha\beta+\beta'\lambda_L}{\alpha'}.
\end{equation}
 
The resolution of the right boundary relation fixes
 \begin{equation} \label{eq:constraint_right}
  \rho = \frac{\mathfrak{c}_N(1-\gamma)(\widetilde\gamma \widetilde\delta-\gamma\delta+\delta\frac{\lambda_{R}}{\mathfrak{a}_N})\delta'}{(\frac{\lambda_{R}}{\mathfrak{a}_N})^2\widetilde\gamma \widetilde\delta}, \qquad 
  \kappa = \frac{\frac{\lambda_{R}}{\mathfrak{a}_N}(\frac{\lambda_{R}}{\mathfrak{a}_N}-\gamma)}{\mathfrak{b}_N(1-\delta)\gamma'}.
 \end{equation}
and imposes the following explicit expression for the right boundary vectors
\begin{equation}
\begin{aligned}
 &\ket{r'_{0}} = \begin{pmatrix}
                1 \\ \frac{\delta \gamma'\mathfrak{a}_N}{\gamma\lambda_{R}} \\ \frac{\delta\lambda_{R}}{\gamma \delta'\mathfrak{a}_N}
               \end{pmatrix},& \quad &
 \ket{r'_{1}} = \begin{pmatrix}
                \frac{1-\gamma}{\gamma} \\ \frac{(1-\delta)\gamma'\mathfrak{a}_N}{\gamma\lambda_{R}} \\ \frac{(1-\delta)\lambda_{R}}{\gamma \delta'\mathfrak{a}_N}
               \end{pmatrix}, \\
 &\ket{r_{00}} = \begin{pmatrix}
               \mathfrak{g}_N \\ \frac{u_{10}^{(N-1)}t_{00}^{(N-1)}\delta(\widetilde\gamma\widetilde\delta-\gamma\delta+\delta\lambda_{R}/\mathfrak{a}_N)\mathfrak{a}_N\mathfrak{g}_N}{\gamma\widetilde\delta\lambda_{R}} \\ \frac{\delta\lambda_{R}}{\mathfrak{f}_N\gamma \delta'}
               \end{pmatrix},& \quad &
 \ket{r_{01}} = \begin{pmatrix}
               \frac{\lambda_{R}-\mathfrak{a}_N\gamma}{y^{(N)}\mathfrak{f}_N\gamma} \\ \frac{u_{10}^{(N-1)}t_{00}^{(N-1)}(\widetilde\gamma\widetilde\delta-\gamma\delta+\delta\lambda_{R}/\mathfrak{a}_N)\mathfrak{a}_N\mathfrak{g}_N}{y^{(N)}\gamma\lambda_{R}} \\ \frac{u_{10}^{(N)}(1-\delta)\lambda_{R}}{\mathfrak{f}_N\gamma \delta'}
               \end{pmatrix}, \\
 &\ket{r_{10}} = \begin{pmatrix}
               \frac{\delta(\lambda_{R}-\mathfrak{a}_N\gamma)}{\mathfrak{f}_N\gamma\widetilde\delta} \\ \mathfrak{g}_N \\
               \frac{\lambda_{R}(\lambda_{R}-\mathfrak{a}_N\gamma)}{\mathfrak{a}_N\mathfrak{f}_Nu_{00}^{(N-1)}t_{10}^{(N-1)}\widetilde\delta \gamma'}
               \end{pmatrix},& \quad &
 \ket{r_{11}} = \begin{pmatrix}
               \frac{\mathfrak{g}_Nz^{(N)}(1-\gamma)}{\gamma} \\ \frac{\mathfrak{g}_Nz^{(N)}(1-\gamma)}{\gamma} \\ 
               \frac{z^{(N)}\lambda_{R}(1-\gamma)(\lambda_{R}-\mathfrak{a}_N\gamma)}{\mathfrak{a}_N\mathfrak{f}_Nu_{00}^{(N-1)}t_{10}^{(N-1)}\gamma\widetilde\delta \gamma'}
               \end{pmatrix}.
\end{aligned}
\end{equation}
 
\subsection{Leading eigenvalue}

Imposing the equality between the two constraints \eqref{eq:constraint_left} and \eqref{eq:constraint_right}, we obtain that the eigenvalue $\lambda(s) = \lambda_{R}\lambda_{L}$ has
 to be the dominant root of a polynomial of order $4$
\begin{equation}
 \lambda^4 - \alpha \gamma \mathfrak{a}_N \lambda^3 - \omega \mathfrak{a}_N^2 \lambda^2 - \beta\delta\xi \mathfrak{a}_N^3 \lambda + \eta \mathfrak{a}_N^4=0
\end{equation}
with 
\begin{equation}
 \begin{aligned}
 \omega &= \mathfrak{b}_N(1-\alpha)(1-\delta)\beta'\gamma'+\mathfrak{c}_N(1-\beta)(1-\gamma)\alpha'\delta' \\
 \xi &= \mathfrak{b}_N\mathfrak{c}_N(1-\alpha)(1-\beta)(1-\gamma)(1-\delta)\frac{\alpha'\beta'\gamma'\delta'}{\widetilde{\alpha}\widetilde{\beta}\widetilde{\gamma}\widetilde{\delta}}\\
 \eta &= (\alpha\beta-\widetilde\alpha\widetilde\beta)(\gamma\delta-\widetilde\gamma\widetilde\delta)\xi.
 \end{aligned}
\end{equation}

 Note that without counting fields, \textit{i.e.} for $s=0$, we have the factorization
 \begin{equation}
  -(\lambda-1)\Big(-\lambda^3+(\alpha\gamma-1)\lambda^2+(\alpha\gamma+\alpha\delta+\beta\gamma-\alpha-\beta-\gamma-\delta+1)\lambda+(\alpha+\beta-1)(\gamma+\delta-1)\Big)=0
 \end{equation}
and we recover that the dominant eigenvalue is $\lambda=1$.

\section{Doob transformation}

\subsection{Explicit form of the Doob transform}
The long-time Doob transform can be defined as \cite{Carollo2017}, 
\begin{equation}
M_{\rm Doob}=\frac{1}{\lambda(s)} \LL M(s) \LL^{-1},
\end{equation}
where $\lambda(s)$ is the leading eigenvalue of the tilted propagator $M(s)$ and $\LL$ is a diagonal operator formed out of components of 
the leading \emph{left} eigenvector $\bm{q}$ of $M(s)$, 
\begin{equation}
\LL_{\bf{n},\bf{n'}}:=\delta_{\bf{n},\bf{n'}} q_{\bf{n}},
\end{equation}
where ${\bf{n}}:=(n_1,n_2,\dots,n_N)$, $n_k\in\{0,1\}$. 
As mentioned in the main text, the final result of this transform and the deformation is only a change to the stochastic boundary matrices $R$ and $L$ that makes the transition rates dependent on configuration of the ingoing and outgoing state in the probability vector. The bulk dynamics remain deterministic. The explicit expression can be written as, 
\begin{equation}
M_{\rm Doob}=M_{\rm o} ^{\rm Doob}M_{\rm e}^{\rm Doob}
\end{equation}
\begin{equation}
 \begin{aligned}
  &  M_{\rm e}^{\rm Doob}= P_{123}P_{345}\cdots P_{N-3,N-2,N-1} R^{\rm Doob}_{N-1,N},\\
  & M_{\rm o} ^{\rm Doob} = P_{234}P_{456} \cdots P_{N-2,N-1,N}L^{\rm Doob}_{12},
 \end{aligned}
\end{equation}
where we defined,
 \begin{alignat}{2}
  & R^{\rm Doob}_{N-1,N}=R^{\rm Doob'}_{N-1,N}F, \qquad &&\left(R^{\rm Doob'}_{N-1,N}\right)_{\bf{n''},\bf{n}}= \left (R_{N-1,N} \right )_{\bf{n''},\bf{n}} (q_{\bm{n}})^{-1} \frac{1}{\lambda(s)},\\
  & L^{\rm Doob}_{12}=L^{\rm Doob'}_{12}G,  \qquad &&\left(L^{\rm Doob'}_{12}\right)_{\bf{n'},\bf{n''}}=q_{\bm{n'}} \left(L_{12}\right)_{\bf{n'},\bf{n''}},
 \end{alignat}
where we used that $[L_{12},P_{234}]=0$
\subsection{Leading left eigenvector}
To calculate the Doob transformation we need the \emph{left} leading eigenvector $\bm{q}M(s)=\lambda(s) \bm{q}$.
In order to reduce the computation of the left eigenvector to a similar problem as the computation of the right eigenvector, and thus 
use partially our previous result, we define,
\begin{equation}
N(s):=F(s)^{-1}M(s)^\intercal F(s)=M_{\rm e}^\intercal G(s) M_{\rm o}^\intercal F(s).
\end{equation}
It is clear that the leading right eigenvector of $N(s)$, $N(s) \bm{o}=\lambda(s) \bm{o}$ is related to $\bm{q}$ via a similarity transformation, 
$\bm{q}=(F(s)\bm{o})^\intercal$. We also have,
\begin{equation}
\begin{aligned}
& M_{\rm e}^\intercal = P_{123}P_{345}\cdots P_{N-3,N-2,N-1} \left(R_{N-1,N}\right)^\intercal \\
& M_{\rm o}^\intercal = \left(L_{12}\right)^\intercal P_{234}P_{456} \cdots P_{N-2,N-1,N}.
\end{aligned}
\end{equation}
Note that $N(s)$ has formally a very similar structure than $M(s)$. The differences are that the odd and even half-time steps are switched, 
$F(s)$ and $G(s)$ are also interchanged and $R_{N-1,N}$ (respectively $L_{12}$) is replaced by $\left(R_{N-1,N}\right)^\intercal$ (respectively $\left(L_{12}\right)^\intercal$).
As previously, we will construct a pair of vector $\bm{o}$ and $\bm{o'}$ satisfying the relations
  $M_{\rm o}^\intercal F(s) \bm{o} = \lambda_{R} \bm{o'}$ and $M_{\rm e}^\intercal G(s) \bm{o'} = \lambda_{L} \bm{o}$.  
Taking $\bm{o}$ (respectively $\bm{o'}$) to be of the same matrix product form as $\bm{p'}$ (respectively $\bm{p}$), 
\begin{equation}
  \begin{aligned}
  & o_{n_1,n_2,\dots,n_N} = \bra{m_{n_1 n_2}} \hat V_{n_3}^{(3)}\hat V_{n_4}^{(4)} \cdots \hat V_{n_{N-2}}^{(N-2)}\hat V_{n_{N-1}}^{(N-1)}\ket{h_{n_N}}\\
  & o'_{n_1,n_2,\dots,N_n} =\bra{m'_{n_1}} \hat W_{n_2}^{(2)}\hat W_{n_3}^{(3)} \cdots \hat W_{n_{N-3}}^{(N-3)}\hat W_{n_{N-2}}^{(N-2)}\ket{h'_{n_{N-1}n_N}}, \label{eq:leftansatz}
  \end{aligned}
 \end{equation}
we arrive to a similar set of equations for the cancellation scheme as we did for $\bm{p'}$ and $\bm{p}$.

The inhomogeneous bulk relations now read (for even site index $j=2i$):
\begin{equation}
\begin{aligned}
 g^{(j-1)}_{nn'} g^{(j)}_{n'n'\!'}\hat W^{(j-1)}_n \hat W^{(j)}_{n'} \hat X^{(j+1)}_{n'\!'} &= 
\hat X^{(j-1)}_{n}\hat V^{(j)}_{\chi(nn'n'\!')} \hat V^{(j+1)}_{n'\!'}, \\
 f^{(j-2)}_{nn'} f^{(j-1)}_{n'n'\!'} \hat X^{(j-2)}_n \hat V^{(j-1)}_{n'} \hat V^{(j)}_{n'\!'} &= \hat W^{(j-2)}_n \hat W^{(j-1)}_{\chi(nn'n'\!')} \hat X^{(j)}_{n'\!'}.
\end{aligned}
\end{equation}
They are similar to \eqref{eq:rel_bulk_comp} but with $f^{(k)}_{nn'}$ and $g^{(k)}_{nn'}$ interchanged. 
An explicit representation of these inhomogeneous bulk relations is thus obtained using our previous results
\begin{equation}
 \hat V^{(j)}_{n} = \left. V^{(j)}_{n} \right|_{\{f^{(k)}_{n'n'\!'}\}\leftrightarrow \{g^{(k)}_{n'n'\!'}\}}, \quad
 \hat W^{(j)}_{n} = \left. W^{(j)}_{n} \right|_{\{f^{(k)}_{n'n'\!'}\}\leftrightarrow \{g^{(k)}_{n'n'\!'}\}}, \quad 
 \hat X^{(j)}_{n} = \left. X^{(j)}_{n} \right|_{\{f^{(k)}_{n'n'\!'}\}\leftrightarrow \{g^{(k)}_{n'n'\!'}\}}
\end{equation}

The boundary equations read
\begin{equation}
\begin{aligned}
 g^{(1)}_{nn'}g^{(2)}_{n'n'\!'} \bra{m_n}\hat W^{(2)}_{n'} \hat X^{(3)}_{n'\!'} &= \bra{m'_{n\chi(nn'n'\!')}}\hat V^{(3)}_{n'\!'},  \\
 \sum_{l,l'=0,1} \left(R^\intercal\right)_{nn'}^{ll'}g^{(N-1)}_{ll'}\ket{h_{ll'}} &= \lambda_{\rm R} \hat X^{(N-1)}_{n}\ket{h'_{n'}}, \\
 \sum_{l,l'=0,1} \left(L^\intercal\right)_{nn'}^{ll'} f^{(1)}_{ll'} \bra{m'_{ll'}} &= \lambda_{\rm L} \bra{m_n}\hat X^{(2)}_{n'},\\
 f^{(N-2)}_{nn'}\!f^{(N-1)}_{n'n'\!'} \hat X^{(N-2)}_{n} \hat V^{(N-1)}_{n'}\!\ket{h'_{n'\!'}} &= \hat W^{(N-2)}_{n}\!\ket{h_{\!\chi(nn'n'\!')n'\!'}}.
\end{aligned}
\end{equation}
They are similar to \eqref{eq:rel_boundaries_comp} but with $f^{(k)}_{nn'}$ and $g^{(k)}_{nn'}$ interchanged and with the matrix
$R$ (respectively $L$) replaced with $R^\intercal$ (respectively $L^\intercal$).
They can be solved explicitly.
The solution for the right boundary equations gives,
\begin{equation}
\rho=\frac{(1-\delta)\gamma'(\widetilde{\gamma} \widetilde{\delta}-\gamma\delta +\delta\frac{\lambda_R }{\mathfrak{a}_N})\mathfrak{b}_N}{(\frac{\lambda_R}{\mathfrak{a}_N})^2\widetilde{\gamma} \widetilde{\delta} }, \qquad 
\kappa=\frac{\frac{\lambda_R}{\mathfrak{a}_N}(\frac{\lambda_R}{\mathfrak{a}_N}-\gamma)}{\mathfrak{c}_N(1-\gamma)\delta'},
\end{equation}
Likewise, we have for the right boundary vectors in the MPS,
  \begin{align}
 & \ket{h_0} = \begin{pmatrix}
     1 \\
     \frac{\delta'\mathfrak{a}_N}{\lambda_R} \\
      \frac{\lambda_R}{\gamma'\mathfrak{a}_N}
      \end{pmatrix}, & \qquad &    \ket{h_1} =    \ket{h_0},  \\
& \ket{h'_{00}} =\begin{pmatrix}
       \mathfrak{f}_N \\
        \frac{t^{(N-1)}_{10} u^{(N-1)}_{00}(\widetilde{\gamma}\widetilde{\delta}- \gamma \delta +\delta \frac{\lambda_R}{\mathfrak{a}_N}) \mathfrak{a}_N \mathfrak{f}_N}{\lambda_R \widetilde{\gamma}}\\
       \frac{\lambda_R}{\mathfrak{g}_N \gamma'} 
        \end{pmatrix}, & \qquad &
 \ket{h'_{01}} =\begin{pmatrix}
         \frac{\lambda_R-\mathfrak{a}_N \gamma}{z^{(N)} \mathfrak{g}_N(1-\gamma)} \\
         \frac{t^{(N-1)}_{10} u^{(N-1)}_{00} (\widetilde{\gamma}\widetilde{\delta}- \gamma \delta +\delta \frac{\lambda_R}{\mathfrak{a}_N}) \mathfrak{a}_N \mathfrak{f}_N}{\lambda_R (1-\gamma)z^{(N)}}\\
          \frac{t^{(N)}_{10}\lambda_R}{\mathfrak{g}_N \gamma'}
           \end{pmatrix},\\
& \ket{h'_{10}} =\begin{pmatrix}
           \frac{\lambda_R-\mathfrak{a}_N\gamma}{\mathfrak{g}_N\widetilde{\gamma}}\\
            \mathfrak{f}_N\\
             \frac{\lambda_R(\lambda_R-\mathfrak{a}_N\gamma)}{\mathfrak{a}_N \mathfrak{g}_N t^{(N-1)}_{00}u^{(N-1)}_{10}\widetilde{\gamma}\delta'}
              \end{pmatrix}, & \qquad &        
\ket{h'_{11}} =\begin{pmatrix}
              \mathfrak{f}_N y^{(N)}\\
              \mathfrak{f}_N y^{(N)}\\
               \frac{y^{(N)} \lambda_R(\lambda_R-\mathfrak{a}_N \gamma)}{\mathfrak{a}_N \mathfrak{g}_N t^{(N-1)}_{00}u^{(N-1)}_{10}\widetilde{\gamma}\delta'}
               \end{pmatrix}.
                    \end{align}

The resolution of the left boundary relation  fixes the value of the parameters $\rho$ and $\kappa$ 
 \begin{equation} 
  \rho = \frac{\lambda_{L}(\lambda_{L}-\alpha)}{(1-\alpha)\beta'}, \qquad 
  \kappa = \frac{(1-\beta)(\widetilde\alpha \widetilde\beta -\alpha\beta+\beta\lambda_{L})\alpha'}{\lambda_{L}^2\widetilde\alpha \widetilde\beta},
 \end{equation}
 and imposes the following explicit expression for the left boundary vectors
 \begin{equation}
  \begin{aligned}
  & \bra{m_{0}}= \Big( 1, \frac{\widetilde\alpha\lambda_L(\zeta-\widetilde\alpha)+\rho(1-\alpha)(\widetilde\alpha\lambda_L-\zeta\alpha')}{u_{00}^{(2)}\widetilde\alpha^2\lambda_L(1-\rho\kappa)} , 
  -\frac{\kappa\widetilde\alpha\lambda_L(\zeta-\widetilde\alpha)+(1-\alpha)(\widetilde\alpha\lambda_L-\zeta\alpha')}{\widetilde\alpha\lambda_L(1-\rho\kappa)(1-\alpha)} \Big) \\
  & \bra{m_{1}}= \bra{m_{0}} \\
  & \bra{m'_{00}}= \Big( \frac{\zeta}{\widetilde\alpha} , -\frac{(1-\alpha)\rho}{y^{(2)}u_{00}^{(2)}\alpha} , \frac{\widetilde\alpha}{\alpha}\Big), \qquad 
   \bra{m'_{01}}= \Big( 0 , \frac{\lambda_L(\zeta-\widetilde\alpha)}{z^{(0)}y^{(2)}u_{00}^{(2)}\widetilde\alpha \beta}, 
   \frac{(\alpha-\lambda_L)(\widetilde\alpha\widetilde\beta-\alpha\beta)\alpha'}{\lambda_L\alpha\widetilde\alpha \beta'} \Big) \\
  & \bra{m'_{10}}= \Big( \frac{\zeta u_{11}^{(1)}}{\widetilde\alpha} , \frac{(1-\alpha)\rho u_{11}^{(1)}}{y^{(2)}u_{00}^{(2)}\widetilde\alpha} , -u_{11}^{(1)} \Big),  \qquad 
   \bra{m'_{11}}= \Big( 0 , 0 , \frac{\zeta \alpha'\widetilde\beta}{\lambda_L(1-\beta)\widetilde\alpha t_{11}^{(1)}} \Big)
  \end{aligned}
 \end{equation}
where we have introduced the parameter
\begin{equation}
\zeta = \frac{\alpha(\widetilde\alpha\widetilde\beta-\alpha\beta+\beta\lambda_L)+\widetilde\alpha \beta\lambda_L}{\alpha\beta'}.
\end{equation}                    
                    
Curiously, the solution for the left eigenvectors $\bm{o}$ is quite similar to the one for the right eigenvectors. 
We note that we could have solved for the left eigenvectors alternatively. 
By first multiplying the first bulk exchange relation \eqref{eq:rel_bulk} by 
$\left(f^{(2k-1)}_{n n'} f^{(2k)}_{n' n''}\right)^{-1}P_{n n' n''}$ and the second one in \eqref{eq:rel_bulk} by 
$\left(g^{(2k-2)}_{n n'} g^{(2k-1)}_{n' n''}\right)^{-1} P_{n n' n''} $ and then replacing $g^{(k)}_{n n'} \to 1/f^{(k)}_{n n'}$
and $f^{(k)}_{n n'} \to 1/g^{(k)}_{n n'}$, we arrive at new bulk exchange relations. These new exchange relations are suitable for 
taking the same ansatz \eqref{eq:leftansatz} and solving $\bm{q}  M_{\rm o} G(s) = \lambda_{R} \bm{q'}$ and $\bm{q'}  M_{\rm e}F(s)  = \lambda_{L} \bm{q}$ directly.
We also would arrive to the same set of boundary equations that may be solved. 
We choose the above because we find that the similarity of the form of the explicit solution to the one for the right eigenvectors is more pronounced.

\end{document}